\documentclass[manuscript]{aastex}

\newcommand{\ltsim}{\protect\raisebox{-0.5ex}{$\:\stackrel{\textstyle <}{\sim}\:$}} 

\shorttitle{Mass-loss evolution of close-in exoplanets}
\shortauthors{Kurokawa and Nakamoto}

\begin{document}


\title{Mass-loss evolution of close-in exoplanets: \\evaporation of hot Jupiters and the effect on population}


\author{H. Kurokawa\altaffilmark{1}}

\email{kurokawa@nagoya-u.jp}

\and

\author{T. Nakamoto\altaffilmark{2}}


\altaffiltext{1}{Department of Physics, Nagoya Univsersity, Furo-cho, Chikusa-ku, Nagoya, Aichi 464-8602, Japan.}
\altaffiltext{2}{Department of Earth and Planetary Sciences, Tokyo Institute of Technology, 2-12-1 Ookayama, Meguro-ku, Tokyo 152-8551, Japan.}


\begin{abstract}
During their evolution, short-period exoplanets may lose envelope mass through atmospheric escape owing to intense XUV (X-ray and extreme ultraviolet) radiation from their host stars. 
Roche-lobe overflow induced by orbital evolution or intense atmospheric escape can also contribute to mass loss.
To study the effects of mass loss on inner planet populations,
we calculate the evolution of hot Jupiters considering mass loss of their envelopes and thermal contraction.
Mass loss is assumed to occur through XUV-driven atmospheric escape and the following Roche-lobe overflow.
The runaway effect of mass loss results in a dichotomy of populations: hot Jupiters that retain their envelopes and super Earths whose envelopes are completely lost. 
Evolution primarily depends on the core masses of planets and only slightly on migration history.
In hot Jupiters with small cores ($\simeq 10$ Earth masses), runaway atmospheric escape followed by Roche-lobe overflow may create sub-Jupiter deserts, 
as observed in both mass and radius distributions of planetary populations.
Comparing our results with formation scenarios and observed exoplanets populations, 
we propose that populations of closely orbiting exoplanets are formed by capturing of planets at/inside the inner edges of protoplanetary disks and subsequent evaporation of sub-Jupiters.

\end{abstract}


\keywords{Planets and satellites: atmospheres - composition - physical evolution - Stars: activity}



\section{Introduction}

To date, hundreds of extrasolar planets and thousands of {\it Kepler} planet candidates have been detected, 
and their statistical properties have been widely investigated \citep[e.g.][]{Howard+2010,Howard+2012}.
Orbital period has been negatively correlated with planetary mass, surface gravity, and mean density \citep[e.g.][]{Mazeh+2005,Southworth+2007,Jackson+2012}.
A negative correlation between orbital period and planetary mass was first reported by \citet{Mazeh+2005}; 
however, only a relatively small number of planets were detected at that time.
A negative correlation between orbital period and surface gravity was later reported by \citet{Southworth+2007}.
These correlations persisted in studies conducted more recently because the number of observed exoplanets increases \citep{Wu+Lithwick2013,Weiss+2013}.
With the detection of inner-orbit super Earths, 
which are clearly separate from the hot Jupiters, 
researchers recognized the so-called \textquotedblleft Desert of sub-Jupiter size exoplanets" at orbital periods of $< {\rm 3\ days}$ in the orbital period-mass diagram \citep{Szabo+Kiss2011,Beauge+Nesvorny2013}. 
This orbital period corresponds to the orbits of $< 0.04\ {\rm AU}$ around solar-mass stars.
The desert has also been found in the orbital period-radius diagram at radii and orbital periods of $3-10\ R_{\rm Earth}$ and $< {\rm 3\ days}$, respectively, 
which include {\it Kepler} planet candidates \citep{Beauge+Nesvorny2013}.

According to the population synthesis of planet formation theory, 
a sub-Jupiter should occur within $\ltsim 1\ {\rm AU}$, 
which originates in gas planets formation at far orbits and planet migration \citep[e.g.][]{Ida+Lin2008,Mordasini+2009,Mordasini+2012}.
However, the statistical analysis of exoplanets revealed that this region is full of planets and the observed desert inside $\ltsim 0.04\ {\rm AU}$ is more compact than predicted \citep{Howard+2010,Howard+2012}.
\citet{Beauge+Nesvorny2013} attributed the compact sub-Jupiter desert to three possible mechanisms.
One of these mechanisms, evaporation of sub-Jupiters by intensive atmospheric escape, is investigated in our study.
The other mechanisms are planet capture at/inside the disk inner edge and interplanetary scattering followed by tidal capture of a planet by its host stars.
These scenarios are compared with the evaporation scenario in Discussion.

\citet{Jackson+2012} showed that orbital period may be correlated with planetary mass if the mass is thermally evaporated by XUV (X-ray and EUV) radiation \citep[see e.g.,][]{Lammer+2003}.
This scenario is applicable to the sub-Jupiter desert as \citet{Beauge+Nesvorny2013} proposed.
\citet{Jackson+2012} adopted a simple energy-limited escape approach \citep{Watson+1981} and assumed a constant radius or density during planetary evolution. 
The energy-limited escape has been widely used in mass loss evolution studies \citep{Lammer+2003,Baraffe+2004,Valencia+2010,Jackson+2010,Leitzinger+2011,Jackson+2012,Kurokawa+Kaltenegger2013}, in which the energy of XUV photons efficiently supplies the escape energy.
However, the models of the upper atmosphere of closely orbiting exoplanets have shown that the escape regime deviates from the energy-limited regime under extremely high XUV \citep{Murray-Clay+2009,Guo2011,Owen+Jackson2012}.
Under these conditions, some of the XUV energy is lost as Lyman-$\alpha$ radiation from excited hydrogen atoms; 
this effect is called \textquotedblleft radiation-recombination limited escape." 
Also, mass loss could cause envelope expansion, 
which further accelerates mass loss \citep{Baraffe+2004,Kurokawa+Kaltenegger2013}.
The evolution of planetary mass and radius must be considered to account for this runaway effect.
Mass loss eventually lead to Roche-lobe overflow, 
in which the envelope is dynamically eroded from the Roche-lobe, 
and a catastrophic evaporation occurs \citep{Kurokawa+Kaltenegger2013}.

In this study, 
we examine how mass loss progresses in inner exoplanets and the consequent effects on exoplanet populations in terms of XUV-driven atmospheric escape and Roche-lobe overflow.
Our model extends the model of \citet{Kurokawa+Kaltenegger2013}, 
which solves the evolution of planetary mass and radius. 
This model enables the effects of runaway mass loss and the Roche-lobe overflow to be observed. 
We newly account for radiation-recombination limited escape by developing a semianalytical model of the upper atmosphere.
Our numerical models are introduced in Section 2.
Section 3 presents our results and compares them with observed exoplanet populations.
Section 4 shows comparisons with previous studies on mass-loss evolution, 
discusses the sensitivity of the results to the model assumptions, 
compares our proposed mechanism with other possible mechanisms of the sub-Jupiter desert, 
and presents some implications for planet formation scenarios.
Section 5 concludes the study.

\section{Numerical models}

The planet is assumed spherically symmetric, with a solar composition envelope and a rock/iron core.
We adopt the Rosseland mean opacity of solar composition gas \citep{Freedman+2008} but use the equation of state (EoS) of hydrogen and helium (${\rm Y = 0.28}$) from the data tables in \citet{Saumon+1995}. 
We introduce two updates into the model of \citet{Kurokawa+Kaltenegger2013}.

The first update introduces the effects of core contraction and its internal heat.
The EoS in the core and the required physical constants are taken from \citet{Wagner+2011}, assuming a rock to iron mass ratio of $0.675:0.325$.
For the upper silicate mantle we use the Vinet EoS \citep{Vinet+1989},
\begin{equation}
p = 3 K_{\rm 0} x^{\frac{2}{3}} (1 - x^{-\frac{1}{3}}) \exp{\Biggl[ \frac{2}{3} (K_{\rm 0}' -1) (1 - x^{-\frac{1}{3}}) \Biggr]}.
\end{equation}
Here $p$ is the pressure, $x = \rho / \rho_{\rm 0}$ is the compression ratio with respect to the ambient density $\rho_{\rm 0}$, 
$K_{\rm 0}$ is the isothermal bulk modulus, and $K_{\rm 0}'$ is the pressure derivative of $K_{\rm 0}$.
The subscript $0$ denotes the ambient conditions.
The higher pressure phases of perovskite, post-perovskite, and iron core, 
are treated by the generalized Rydberg EoS \citep{Wagner+2011}, 
\begin{equation}
p = 3 K_{\rm 0} x^{K_{\infty}'} (1 - x^{-\frac{1}{3}}) \exp{\Biggl[ \biggl(\frac{3}{2} K_{\rm 0}' - 3 K_{\infty}' +\frac{1}{2}\biggr) \biggl(1 - x^{-\frac{1}{3}}\biggr) \Biggr]},
\end{equation}
where the subscript $\infty$ denotes the limit of infinitely large pressure.
To incorporate the contribution of the core heat, Eq. 9 in \citet{Kurokawa+Kaltenegger2013} is substituted by the energy conservation equation \citep[e.g.][]{Lopez+2012},
\begin{equation}
\int_{M_{\rm core}}^{M_{\rm p}} T \frac{{\rm d} S}{{\rm d} t} {\rm d} M_r = L_{\rm int} - L_{\rm radio} - C_p \frac{{\rm d} T_{\rm core}}{{\rm d} t}. \label{EnergyConservation}
\end{equation}
On the left hand side of Eq. \ref{EnergyConservation}, 
$M_{\rm p}$, $M_{\rm core}$, and $M_r$ denote the planetary mass, the core mass, and the enclosed mass at distance $r$, respectively.
And $T$ is the temperature, $S$ is the entropy in the convective layer of the envelope, and $t$ is the time.
On the right hand side, $R_{\rm p}$ and $L_{\rm int}$ denote the radius and the intrinsic luminosity of the planet, respectively, 
$L_{\rm radio}$ is the rate of heat production by the radioactive elements in the rocky layer, 
$C_p$ is the heat capacity of the core, and $T_{\rm core}$ is the temperature at the top of the core.
The abundance of radioactive elements to calculate $L_{\rm radio}$ and the value of $C_p$ are assumed to be the same as those in \citet{Yukutake2000}.

The other update is the mass loss due to atmospheric escape.
To accommodate this loss, 
we consider the transition from the energy-limited escape to the radiation-recombination limited regime.
The rate of mass loss is calculated for each regime and the regime yielding the smaller rate is assumed.
This scheme smoothly connects the radiation-recombination escape induced by high XUV to the energy-limited escape under lower XUV conditions.
The Roche-lobe overflow induced by the atmospheric escape is calculated as described in \citet{Kurokawa+Kaltenegger2013}.

Under lower XUV conditions, 
the energy-limited escape is modeled by the formula of \citet{Lopez+2012}, given by,
\begin{equation}
\frac{{\rm d} M_{\rm p}}{{\rm d} t} = \frac{\eta \pi F_{\rm XUV} R_{\rm XUV}^3}{G M_{\rm p} K_{\rm tide}},
\end{equation}
where $\eta$ is the efficiency, $F_{\rm XUV}$ is the incoming XUV energy flux, $R_{\rm XUV}$ is the radius of XUV photon absorption, $G$ is the gravitational constant, and $K_{\rm tide}$ is a tidal correction factor.
Of these parameters, only the XUV radius $R_{\rm XUV}$ is calculated differently from \citet{Kurokawa+Kaltenegger2013}, which assumed a thin layer between $R_{\rm p}$ and $R_{\rm XUV}$. 
Following \citet{Lopez+2012}, we calculated an isothermal structure from $R_{\rm p}$ to $R_{\rm XUV}$ and defined $R_{\rm XUV}$ at the $1\ {\rm nbar}$ level.
We neglected photochemical processes, which complicate the temperature structure in the upper atmosphere. 
This procedure yielded a rough estimate of the XUV radius $R_{\rm XUV}$.
The influence is evaluated in Discussion.

\begin{figure}
\begin{center}
\includegraphics[scale=1.0]{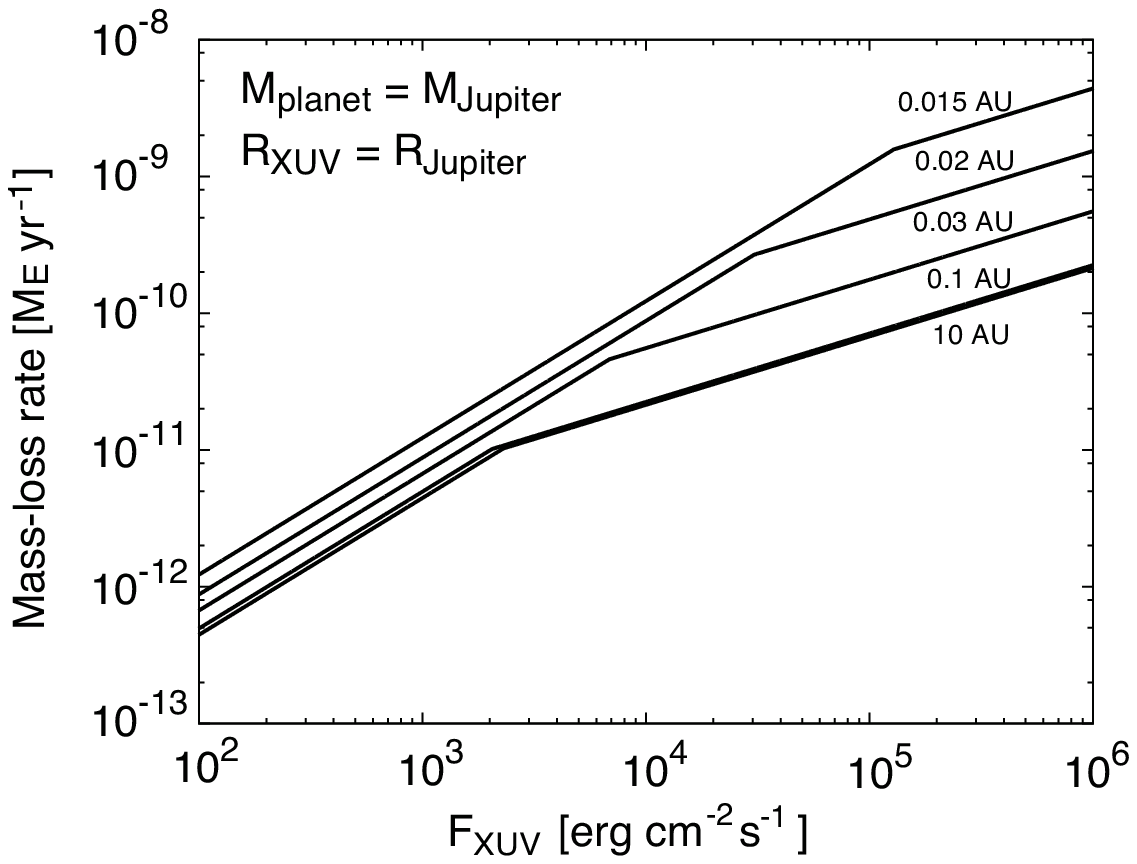}
\caption{Rate of mass loss by thermal atmospheric escape. 
The mass and radius of the planets are $M_{\rm p} = 1\ M_{\rm Jupiter}$ and $R_{\rm XUV} = 1\ R_{\rm Jupiter}$, respectively. Results are plotted at semimajor axes of $0.015\ {\rm AU}$, $0.02\ {\rm AU}$, $0.03\ {\rm AU}$, $0.1\ {\rm AU}$, and $10\ {\rm AU}$ (ordered from top to bottom). 
\label{Fxuv-Mdot}}
\end{center}
\end{figure}

\begin{figure}
\begin{center}
\includegraphics[scale=1.0]{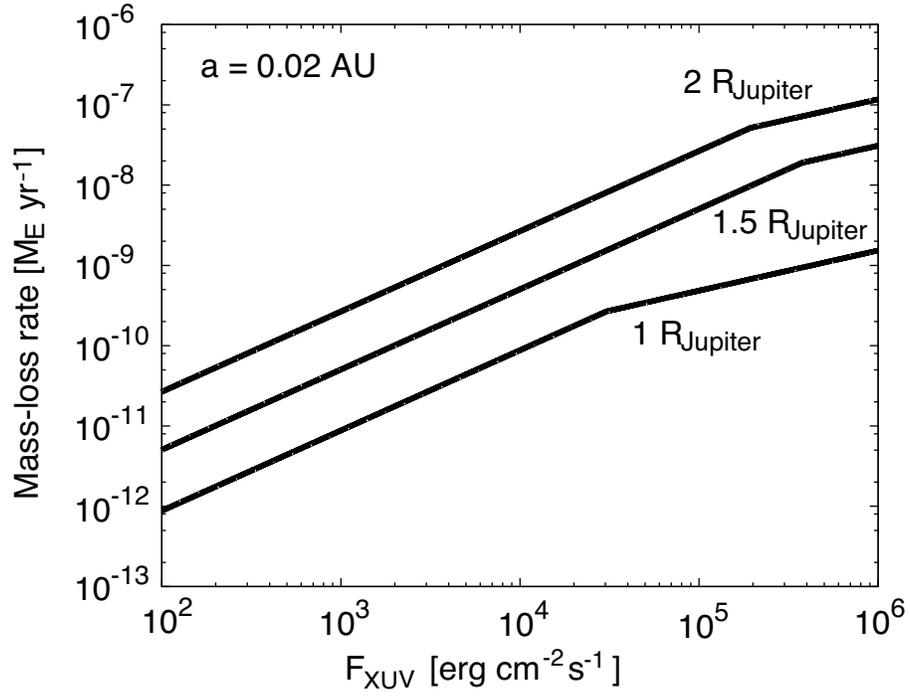}
\caption{Rate of mass loss by thermal atmospheric escape. 
The planetary mass is $M_{\rm p} = 1\ M_{\rm Jupiter}$. 
Results are plotted for planetary radii $R_{\rm planet} = 2,\ 1.5,\ {\rm and}\ 1\ R_{\rm Jupiter}$ (ordered from top to bottom). 
The semi-major axis of the planets is $0.02\ {\rm AU}$. 
\label{Fxuv-Mdot_R}}
\end{center}
\end{figure}

Higher XUV conditions are treated in a semianalytical model of the radiation-recombination limited escape. 
The model is based on the analytical approach of \citet{Murray-Clay+2009}, which accounts for tidal and nonhydrostatic effects.
In the radiation-recombination limited regime, the hydrogen atmosphere above $R_{\rm XUV}$ is almost fully ionized by XUV radiation and holds $\sim 10^4\ {\rm K}$ by Lyman-$\alpha$ radiation emitted from excited hydrogen atoms.
Here we model the regime as an isothermal transsonic flow.
The rate of mass loss at the sonic point is given by,
\begin{equation}
\frac{{\rm d} M_{\rm p}}{{\rm d} t} = 4 \pi \rho_{\rm s} w_{\rm s} r_{\rm s}^2, \label{Mdot_sonic}
\end{equation}
where the subscript s denotes the sonic point and $w_{\rm s}$ is the upward velocity which is equal to the isothermal sound speed $c_T$ of ionized hydrogen gas at $10^4\ {\rm K}$.
The radius at the sonic point $r_{\rm s}$ is calculated as,
\begin{equation}
2 c_T^2 = \frac{G M_{\rm p}}{r_{\rm s}} - 3\frac{G M_{\rm star} r_{\rm s}^2}{a^3}, \label{sonic}
\end{equation}
where the second term in the right hand side represents the tidal contribution. 
Here $M_{\rm star}$ is the mass of the host star and $a$ is the orbital radius.
The density at the sonic point is obtained from the isothermal structure equation,
\begin{equation}
\rho_{\rm s} = \rho_{\rm base} \exp \Biggl[ -\frac{G M_{\rm star}}{c_T^2}(r_{\rm base}^{-1} - r_{s}^{-1}) -\frac{1}{2} + \frac{1}{2} \bigg(\frac{w_{\rm base}}{c_T}\bigg)^2 
+ \frac{3 G M_{\rm star}}{2 a^3 c_T^2} (r_{\rm s}^2 - r_{\rm base}^2)\Biggr], \label{rho_sonic}
\end{equation}
where the subscript \textquotedblleft base" denotes the base of the flow, which corresponds to $R_{\rm XUV}$.
The density at the base is the balance between photoionization and radiative recombination \citep{Murray-Clay+2009},
\begin{equation}
\frac{F_{\rm XUV}}{h \nu_0} \sigma_{\nu_0} n_{\rm 0,base} = n^2_{\rm +,base} \alpha_{\rm rec}, \label{rho+_base}
\end{equation}
where $h \nu_0$ is photon energy ($\sim 20\ {\rm eV}$), $\sigma_{\nu_0}$ is the cross section for photoionization of hydrogen, $n_0$ and $n_+$ are the number densities of neutral and ionized hydrogen, respectively, and $\alpha_{\rm rec} = 2.7 \times 10^{-13}\ {\rm cm^3\ s^{-1}}$ is the radiative recombination coefficient for hydrogen ions. 
The neutral number density at the base $n_{\rm 0,base}$ is estimated by,
\begin{equation}
n_{\rm 0,base} \sim \frac{1}{\sigma_{\nu_0} H_{\rm base}} \sim \frac{m_{\rm H} g_{\rm base}}{2 \sigma_{\nu_0} k_B T}, \label{rho0_base}
\end{equation}
where $H$ is the scale height, $g$ is the gravitational acceleration, and $k_B$ is the Boltzmann constant.
Eqs. \ref{Mdot_sonic}-\ref{rho0_base} provide the rate of mass loss in the radiation-recombination regime.
If the semianalitical model places the sonic point $r_{\rm s}$ below the wind base $R_{\rm XUV}$, 
we assume that $r_{\rm s} = R_{\rm XUV}$.
This situation occurs for highly inflated planets.
Modeling the escape flow in such scenarios is beyond the scope of this study. 
The resulting rate of mass loss is plotted as a function of incoming XUV flux is shown in Figs. \ref{Fxuv-Mdot} and \ref{Fxuv-Mdot_R}.
As the incoming XUV flux decreases, the mass-loss regime smoothly changes from the radiation-recombination limited escape, in which rate of mass loss is proportional to $F_{\rm XUV}^{\frac{1}{2}}$, to the energy-limited regime, where in rate of mass loss is proportional to $F_{\rm XUV}$.
Inflated planets lose mass at a faster rate because their upper atmospheres are deficiently bound.
In Section 3, we will show that our results are strongly affected by this property.

Because most of the observed exoplanets orbit G-type stars, 
we selected the Sun as the model host star.
As in \citet{Kurokawa+Kaltenegger2013}, we adopted the XUV evolution model of \citet{Ribas+2005}, 
which is based on observations of nearby solar analogs and which assumes a typical saturation phase of $0.1\ {\rm Gyr}$ \citep{Jackson+2012}.

The thermal evolution is modeled from the start time $t_{\rm age} = 0\ {\rm yr}$, 
while mass loss starts at $t_{\rm age} = 10^7\ {\rm yr}$. 
These timings are based on the observed lifetime of protoplanetary disk gases, 
namely a few million years \citep[e.g.][]{Haisch+2001}.
The initial entropy in the planetary convective layer ($9.2\ k_{\rm B}\ {\rm baryon^{-1}}$) was estimated from \citet{Marley+2007}'s model of a $1\ M_{\rm Jupiter}$ planet.
If the planet yields no hydrostatic solution or if $R_{\rm XUV}$ becomes larger than the initial $R_{\rm rl}$, 
the planet initially exists in a Roche-lobe overflow state. 
In this situation, which typifies small-core planets of $M_{\rm p}\ \ltsim 100\ M_{\rm Earth}$, 
the initial condition is set to $R_{\rm XUV} = R_{\rm rl}$.
We calculate the evolution of migrated planets as well as planets formed {\it in situ}.
Migrated planets that formed in outer regions cool more rapidly and possess lower entropy than their {\it in situ} counterparts. 
To capture their migration history, we assign a different initial entropy to migrated planets.
On the basis of thermal evolution calculations, 
the initial entropy of planets above and below $200\ M_{\rm Earth}$ is assumed to be $8.5\ k_{\rm B}\ {\rm baryon^{-1}}$ and $8\ k_{\rm B}\ {\rm baryon^{-1}}$, respectively (see Section 3 for details).

\section{Results}

\subsection{Properties of mass-loss evolution}

\begin{figure}
\begin{center}
\includegraphics[scale=0.5]{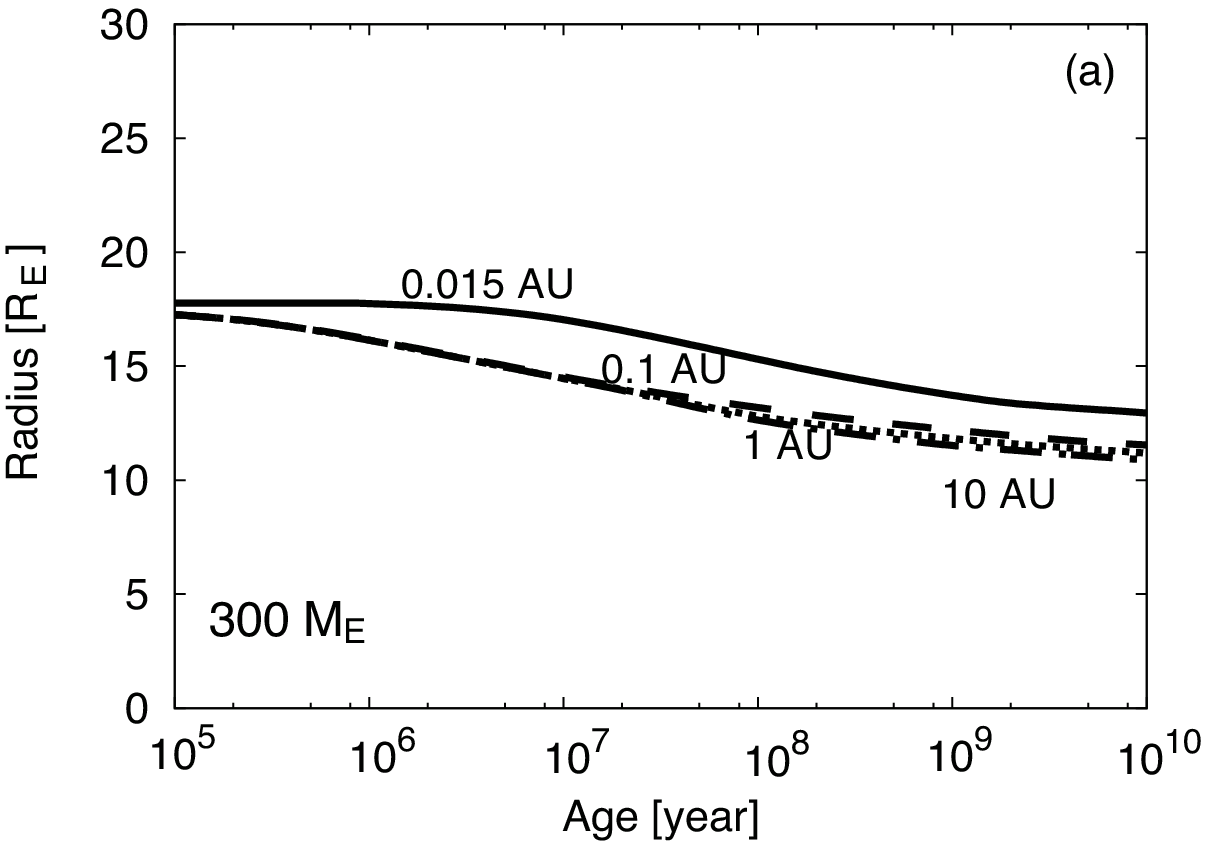}
\includegraphics[scale=0.5]{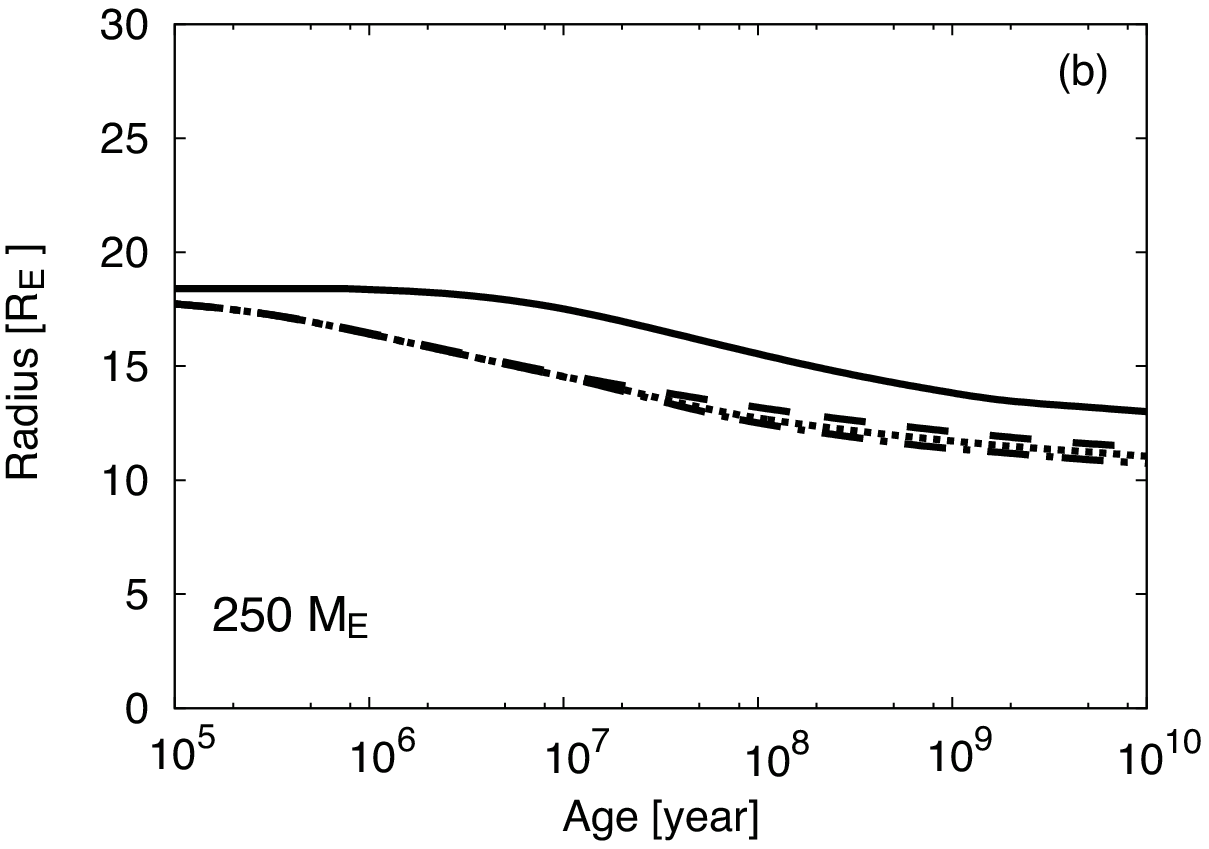}
\includegraphics[scale=0.5]{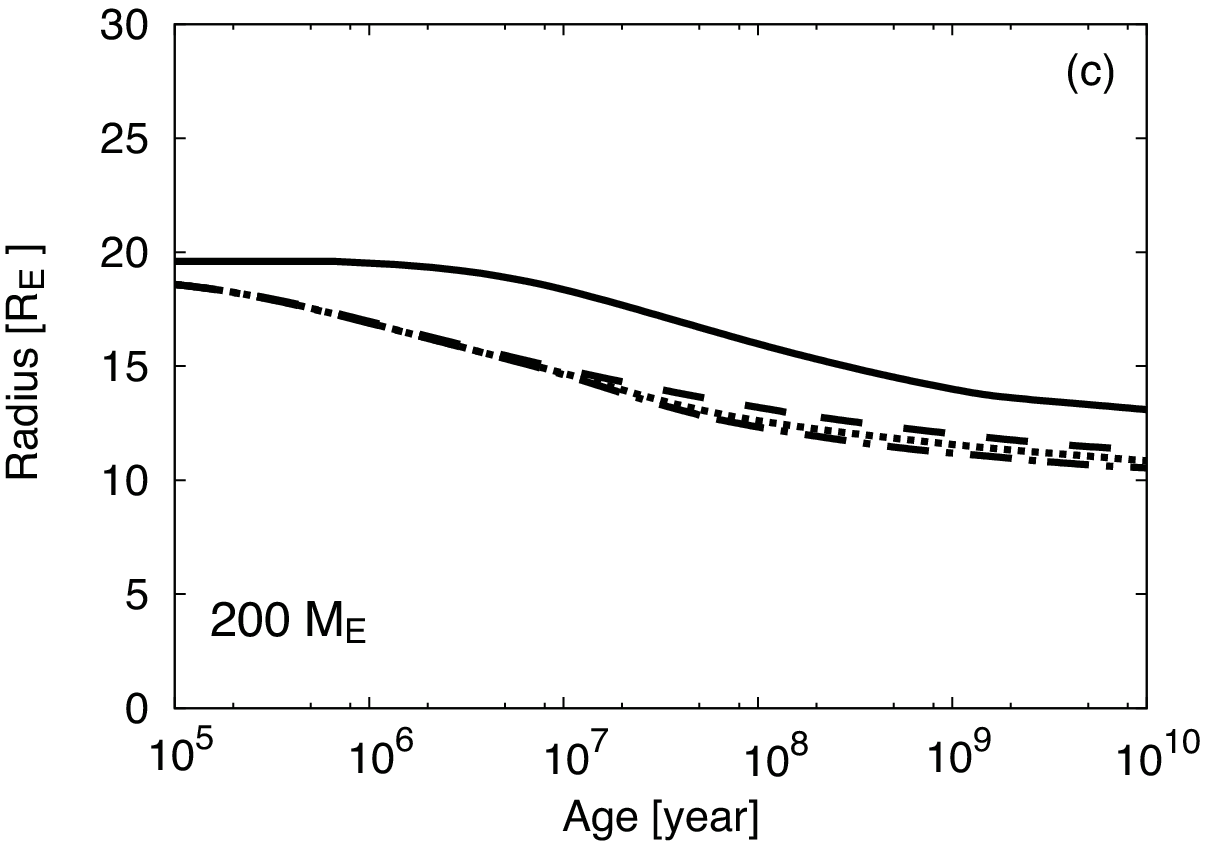}
\includegraphics[scale=0.5]{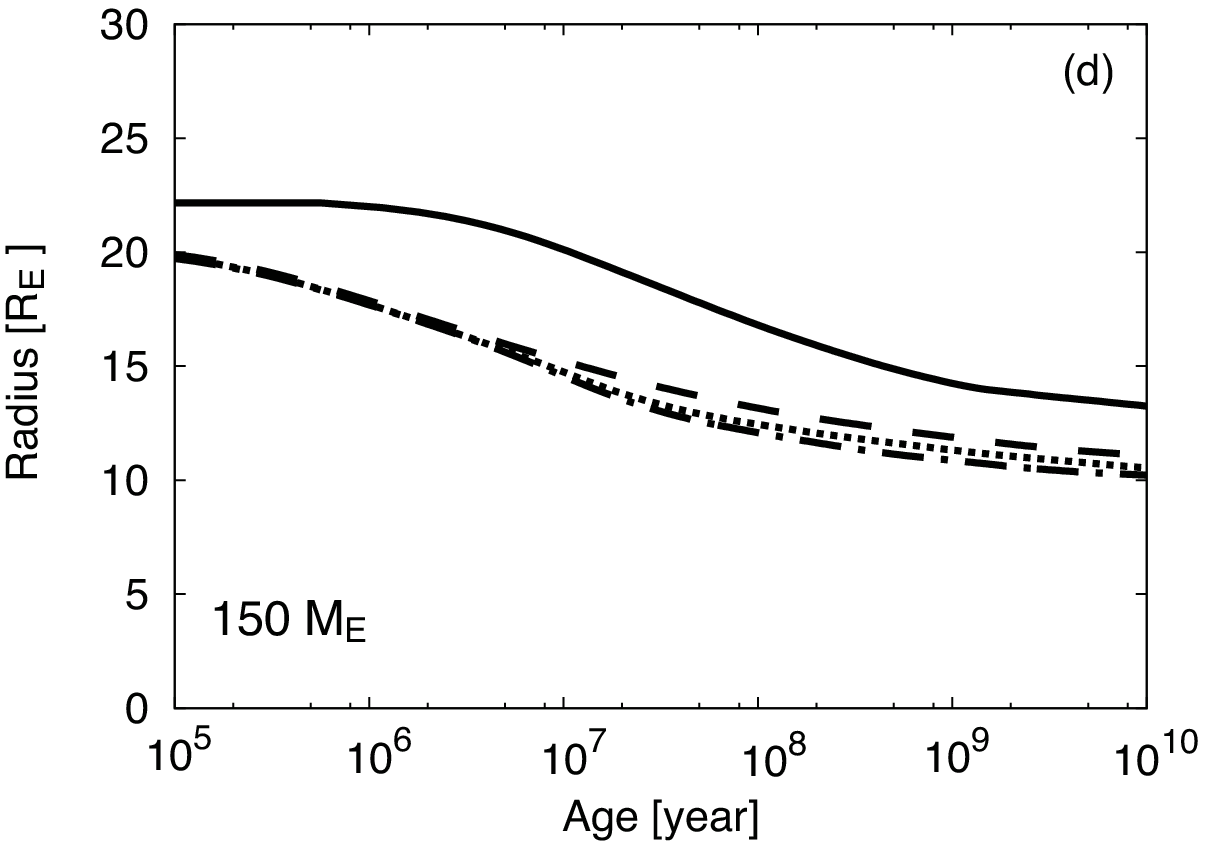}
\includegraphics[scale=0.5]{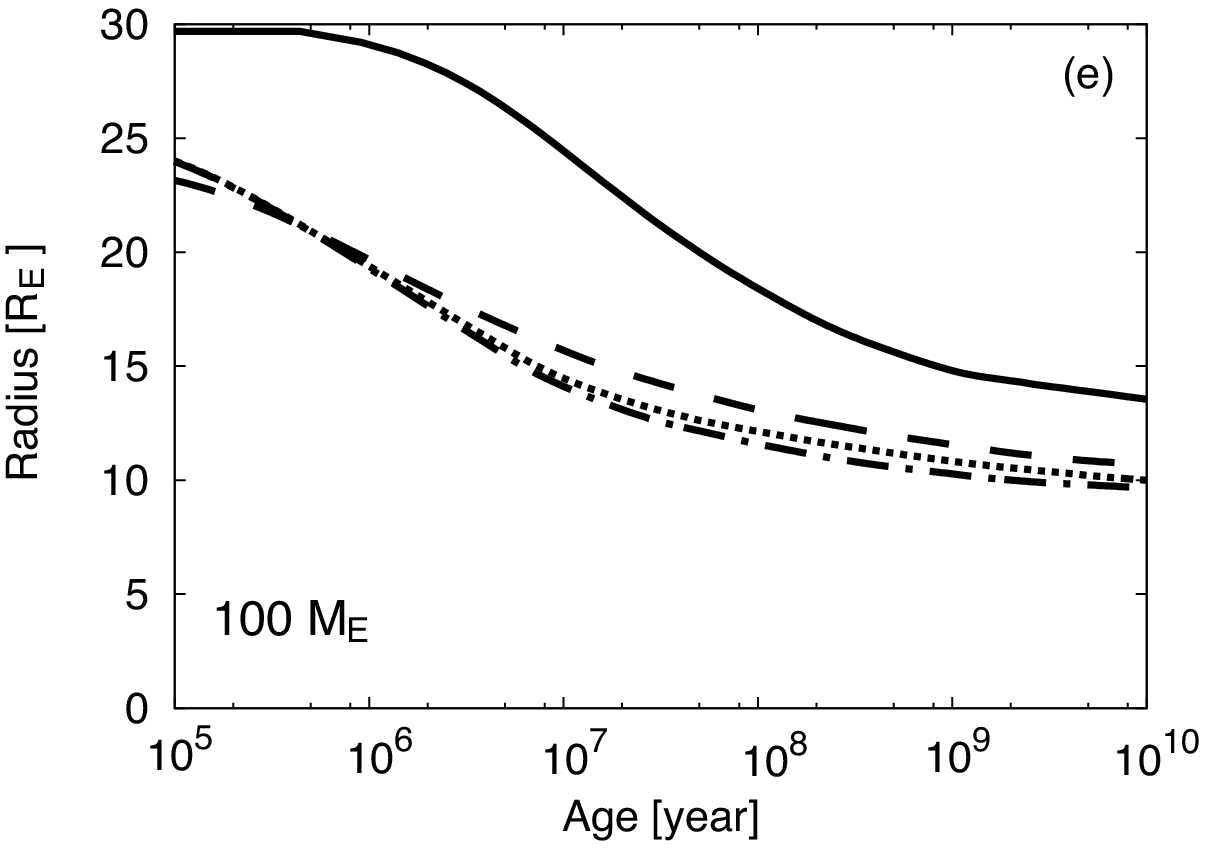}
\includegraphics[scale=0.5]{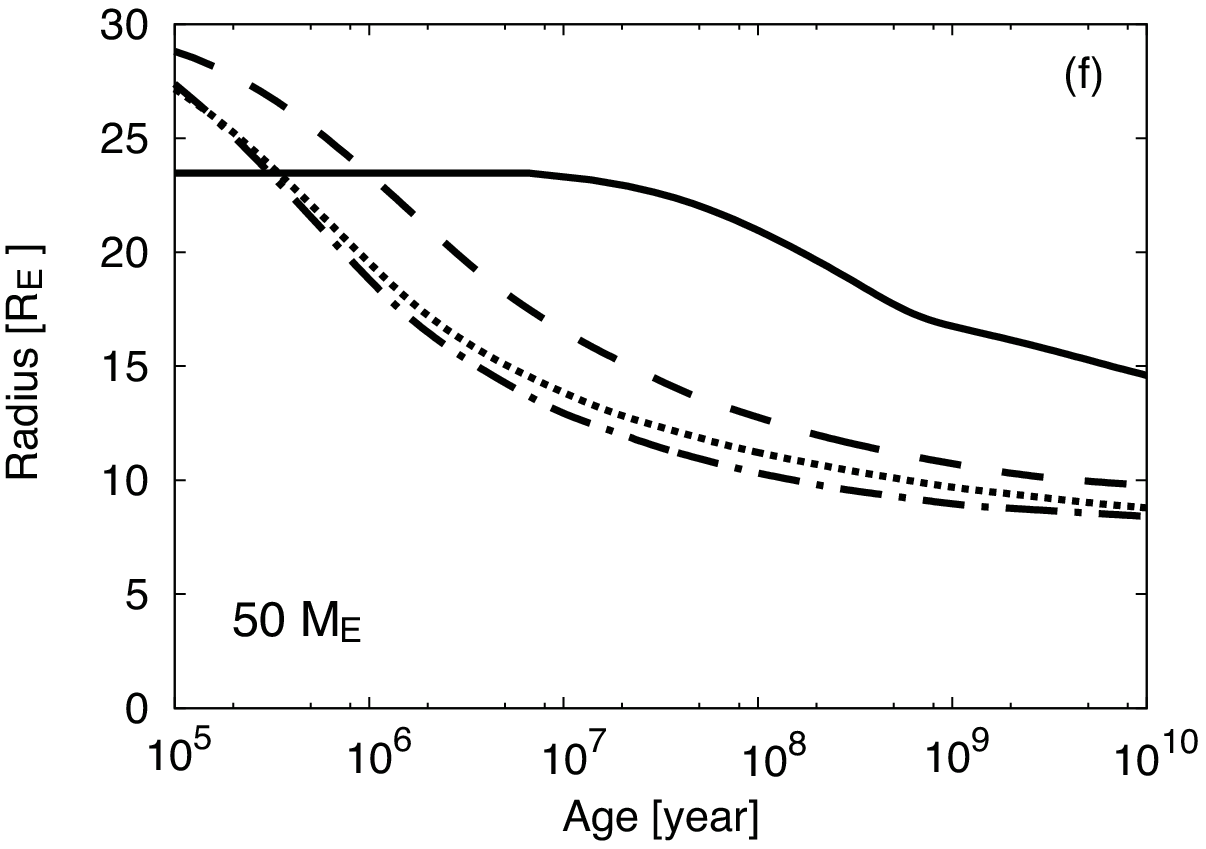}
\caption{Evolution of the radius of gas planets assuming no mass loss. 
All core masses are $10\ M_{\rm Earth}$. Planetary masses are (a) $300\ M_{\rm Earth}$, (b) $250\ M_{\rm Earth}$, (c) $200\ M_{\rm Earth}$, (d) $150\ M_{\rm Earth}$, (e) $100\ M_{\rm Earth}$, and (f) $50\ M_{\rm Earth}$. 
Planetary orbits are $0.015\ {\rm AU}$ (solid line), $0.1\ {\rm AU}$ (dashed line), $1\ {\rm AU}$ (dotted line), and $10\ {\rm AU}$ (dash-dotted line). 
Initial entropies are $9.2\ k_{\rm B}\ {\rm baryon^{-1}}$ except for the $50\ M_{\rm Earth}$ planet at $ 0.015\ {\rm AU}$, for which no hydrostatic solution exists for the initial entropy. 
In this case, a lower initial entropy that admits a hydrostatic solution is assumed. \label{time-radius_10Mcore}}
\end{center}
\end{figure}

\begin{figure}
\begin{center}
\includegraphics[scale=0.5]{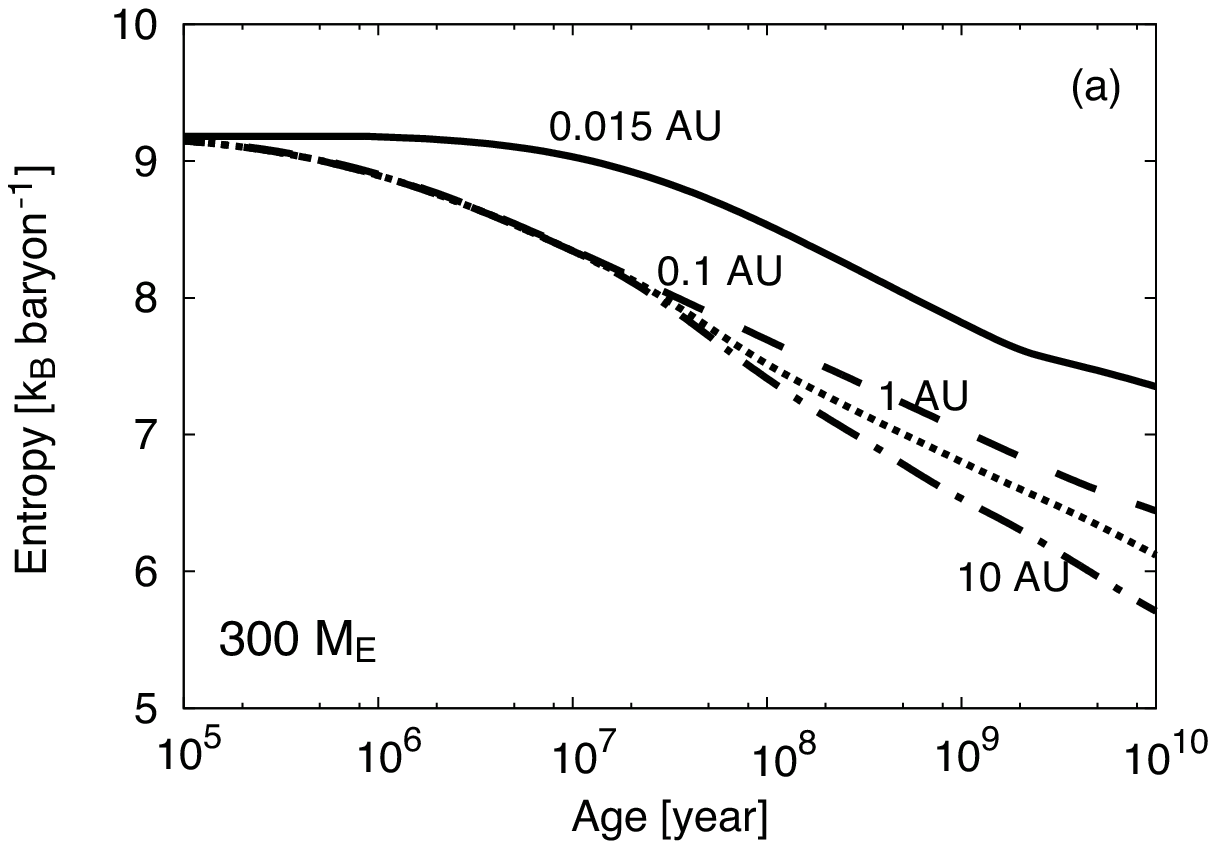}
\includegraphics[scale=0.5]{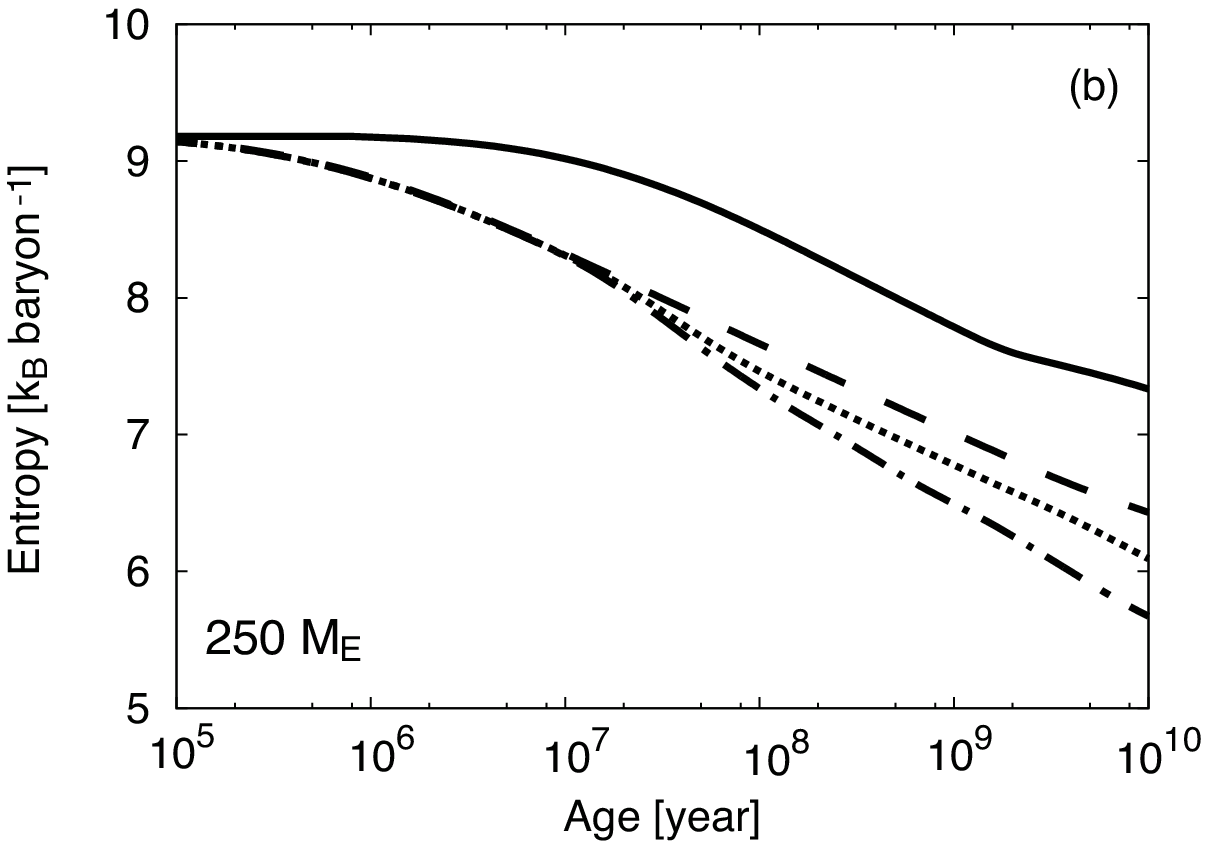}
\includegraphics[scale=0.5]{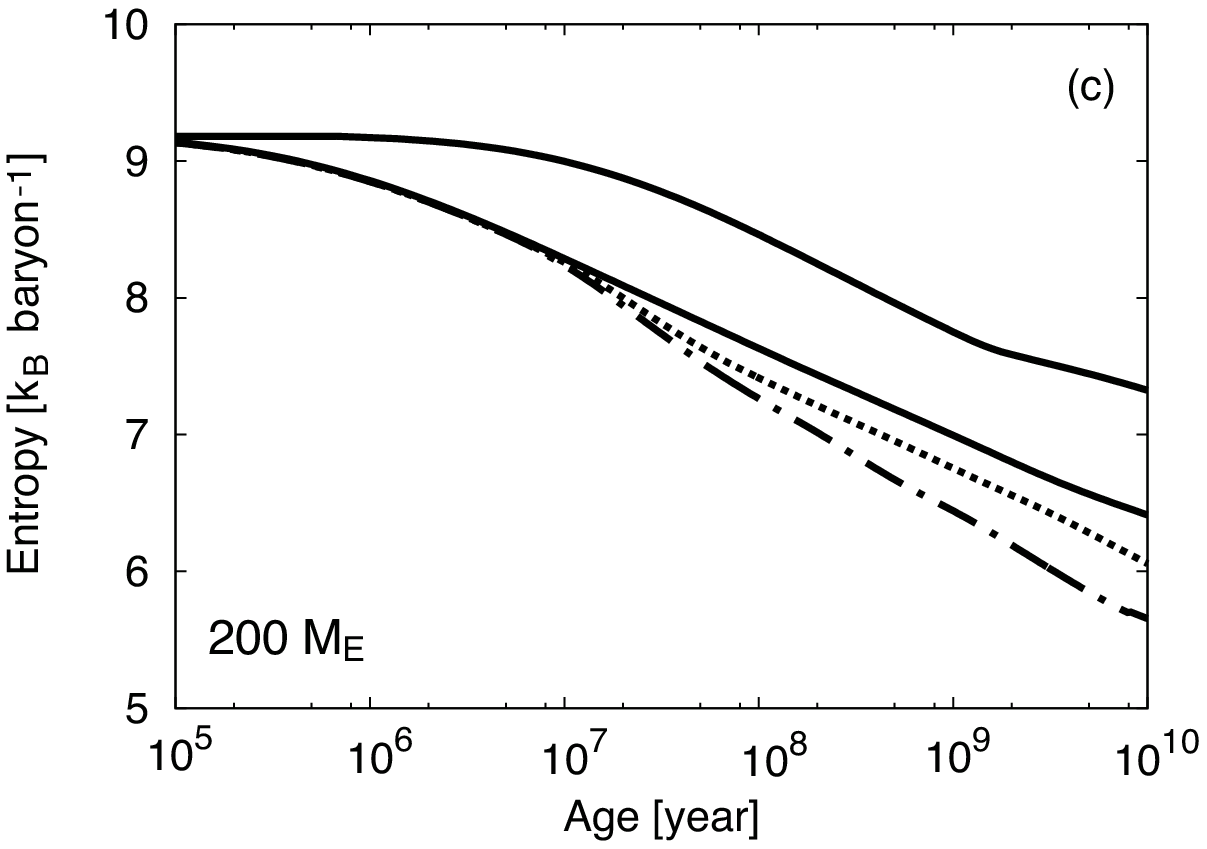}
\includegraphics[scale=0.5]{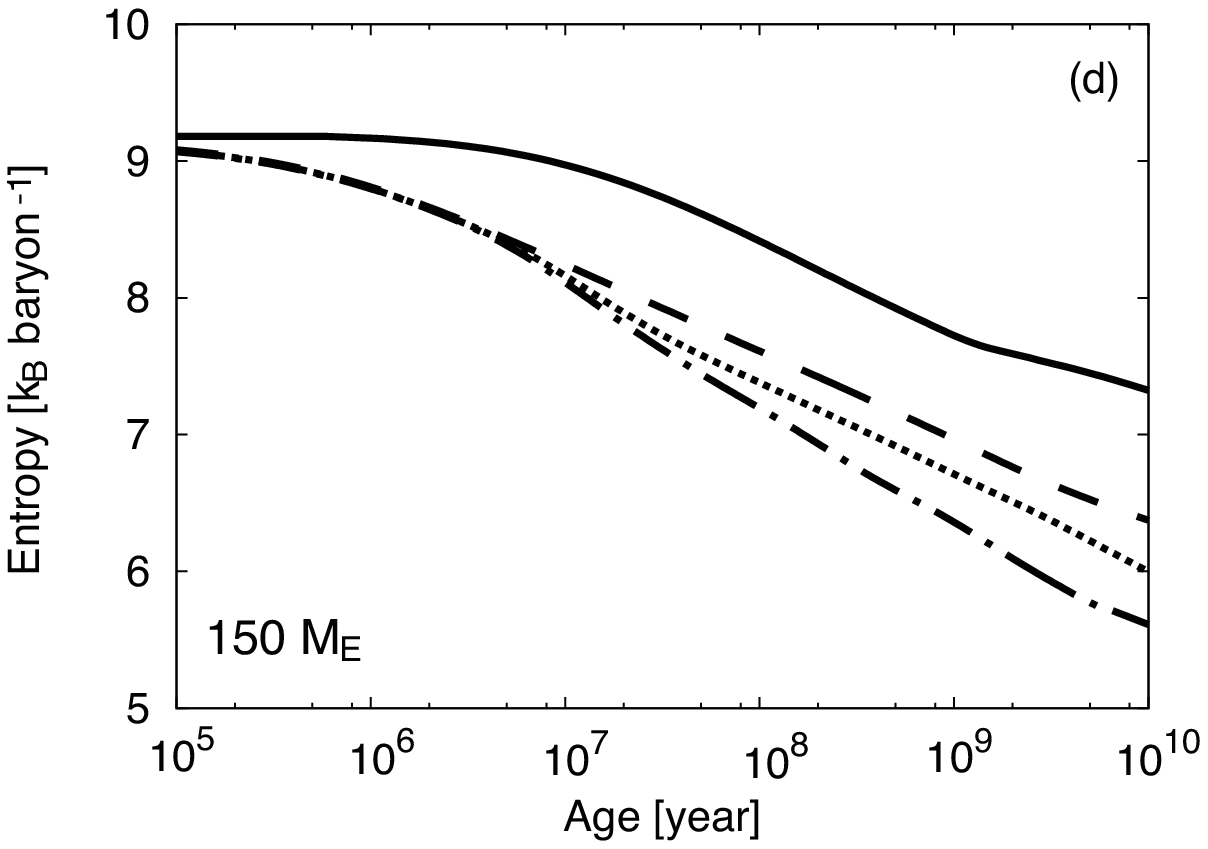}
\includegraphics[scale=0.5]{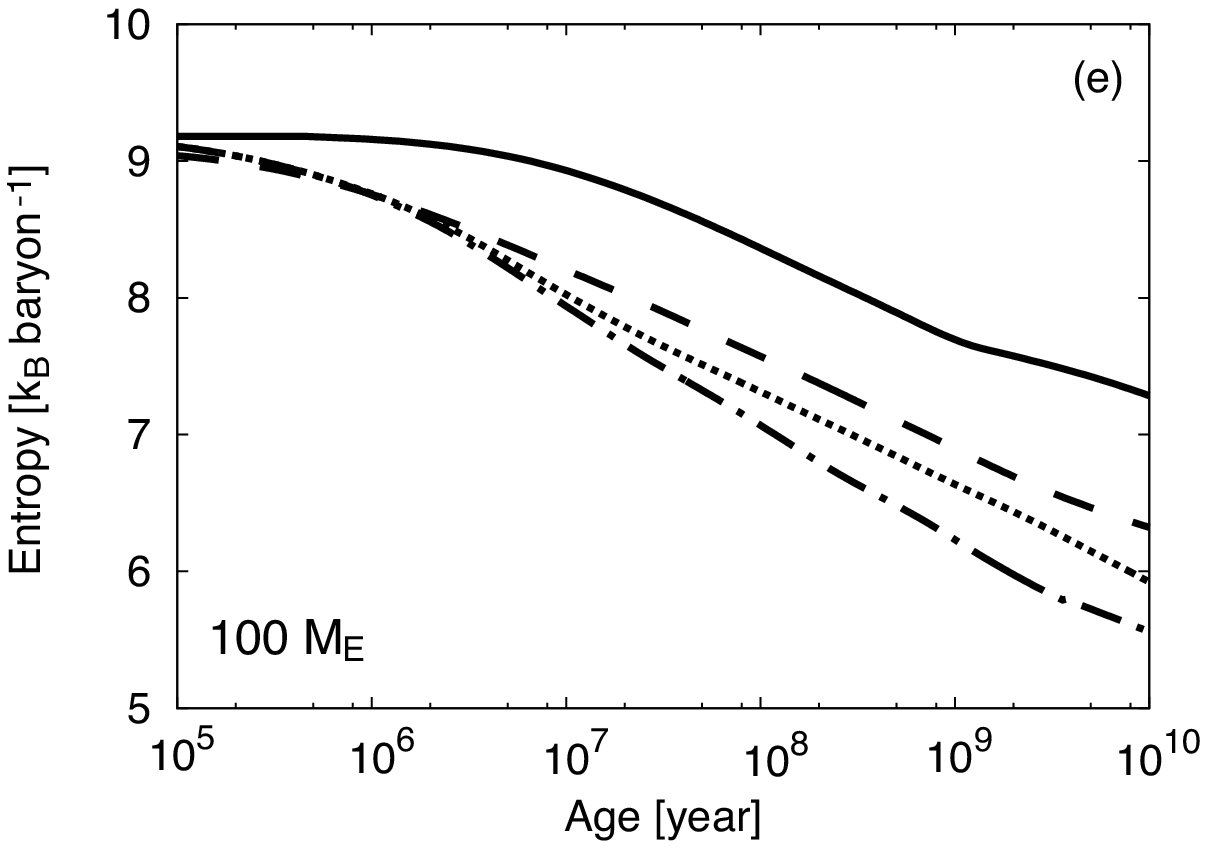}
\includegraphics[scale=0.5]{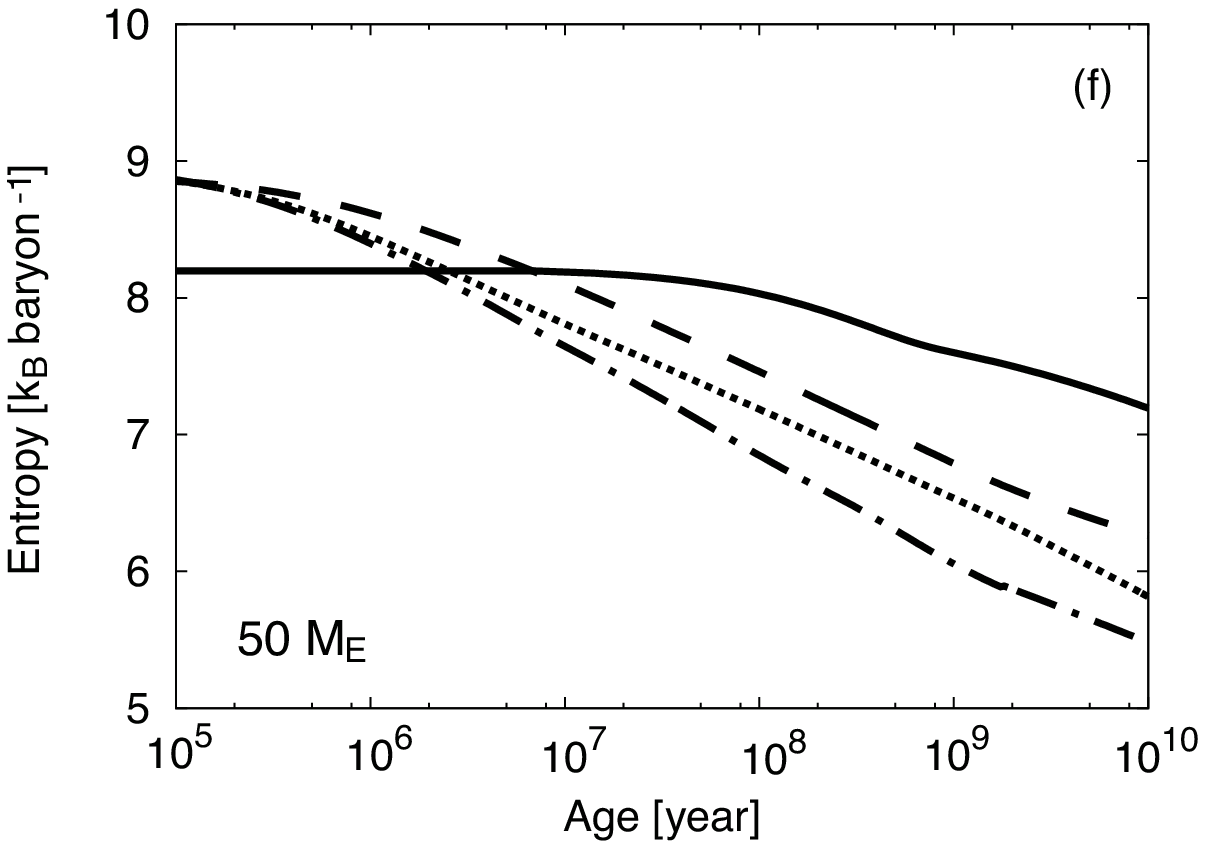}
\caption{Evolution of the entropy of gas planets assuming no mass loss. 
Core masses are $10\ M_{\rm Earth}$. 
Planetary masses are (a) $300\ M_{\rm Earth}$, (b) $250\ M_{\rm Earth}$, (c) $200\ M_{\rm Earth}$, (d) $150\ M_{\rm Earth}$, (e) $100\ M_{\rm Earth}$, and (f) $50\ M_{\rm Earth}$. 
Planetary orbits are $0.015\ {\rm AU}$ (solid line), $0.1\ {\rm AU}$ (dashed line), $1\ {\rm AU}$ (dotted line), and $10\ {\rm AU}$ (dash-dotted line). Initial entropies are $9.2\ k_{\rm B}\ {\rm baryon^{-1}}$ except for the $50\ M_{\rm Earth}$ planet at $ 0.015\ {\rm AU}$, for which no hydrostatic solution exists for the initial entropy. 
In this case, lower initial entropy that admits a hydrostatic solution is assumed. \label{time-entropy_10Mcore}}
\end{center}
\end{figure}

To observe the effects of stellar irradiation on thermal evolution, 
which is related to planetary migration history, 
we first simulated the thermal evolution without mass loss at different orbital radii. 
The results are presented in Figs. \ref{time-radius_10Mcore} and \ref{time-entropy_10Mcore}. 
As planets cool, their large initial radii gradually shrink. 
Planets orbiting closer to their parent star maintain larger radii because they receive heat from stellar radiation, 
which in turn reduces their cooling rate. 
In addition, the radii of low mass planets are more susceptible to cooling effects than those of heavier planets.
Small-mass planets are highly inflated in their nascent stages, i.e., when they are hot, 
which affects their mass-loss evolution as shown later. 
The initial entropy in the migrated model can be estimated from Fig. \ref{time-entropy_10Mcore}. 
Planetary migration induced by interaction with protoplanetary disk gas should terminate when the disk dissipates. 
Assuming that migrated planets form at distant orbits ($\sim 1-10\ {\rm AU}$) and migrated $10^7\ {\rm yr}$ years later, 
the initial entropy of migrated planets above and below $200\ M_{\rm Earth}$ is estimated as $8.5\ k_{\rm B}\ {\rm baryon^{-1}}$ and $8\ k_{\rm B}\ {\rm baryon^{-1}}$, respectively.

\begin{figure}
\begin{center}
\includegraphics[scale=1.0]{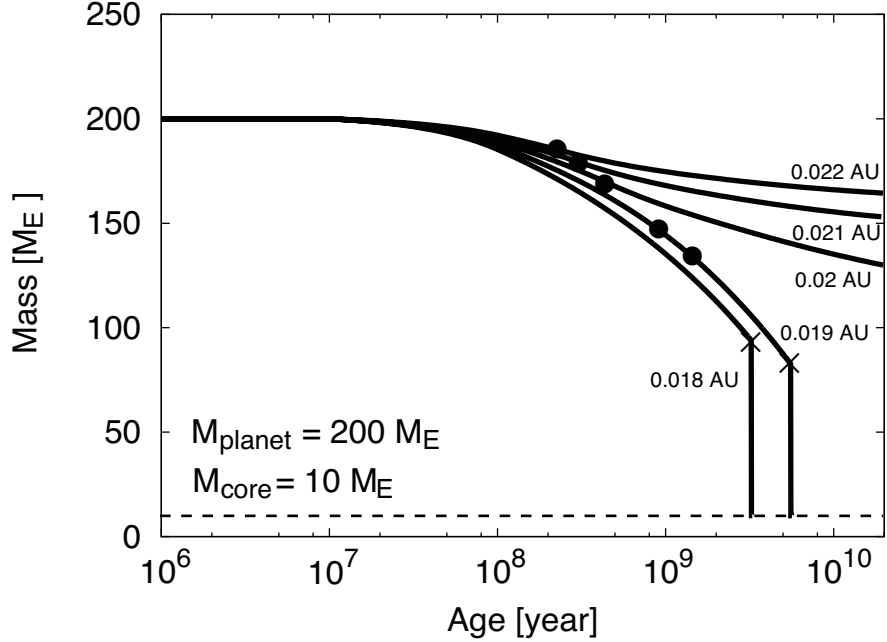}
\caption{Mass evolution of $200\ M_{\rm Earth}$ planets as functions of time. 
The semi-major axes range from $0.018\ {\rm AU}$ to $0.022\ {\rm AU}$ in $0.001\ {\rm AU}$ increments. 
All core masses are $10\ M_{\rm Earth}$ (indicated by the dashed line). 
Filled circles indicate the changeover times of the atmospheric escape regime; from radiation-recombination limited to energy-limited.
The second circle at $0.019\ {\rm AU}$ marks the transition from the energy-limited regime to the radiation-recombination limited regime.
Crosses indicate Roche-lobe overflow events. \label{Evo_RRL_10Mcore300ME_0015-0019AU}}
\end{center}
\end{figure}

The mass-loss evolution and the effect of the orbital radius are shown in Fig. \ref{Evo_RRL_10Mcore300ME_0015-0019AU}.
The planets are $200\ M_{\rm Earth}$ with core masses of $10\ M_{\rm Earth}$. 
Planets of semimajor axis below $0.019 {\rm AU}$ are completely evaporated within $10\ {\rm Gyr}$. 
Although we account for radiation-recombination limited escape, 
which weakly depends on XUV insolation ($F_{\rm XUV}^{0.5}$) and which reduces the mass-loss rate, 
complete evaporation is possible. 
Models in which the atmosphere escapes solely by energy-limited thermal escape also permit total evaporation \citep{Baraffe+2004,Valencia+2010,Jackson+2010,Kurokawa+Kaltenegger2013}. 
In the event of complete evaporation, 
the planets lose large quantities of their envelopes in the radiation-recombination limited regime (Fig. \ref{Evo_RRL_10Mcore300ME_0015-0019AU}).
Once a planet enters the energy-limited regime, it loses mass at a much slower rate.
Because this regime strongly depends on the XUV flux ($F_{\rm XUV}^{1}$), 
the temporal decline of the stellar XUV radiation markedly affects the rate of mass loss.
Besides the thermal atmospheric escape, Roche-lobe overflow contributes to the complete evaporation.
Once Roche-lobe overflow occurs, most of the planet's envelope is lost.
The remnant (a thin envelope of mass $< 1\ M_{\rm Earth}$) dissipates by thermal atmospheric escape over a short time scale.
Planets with slightly larger semimajor axis (in this case, $> 0.02\ {\rm AU}$) retain most of their envelopes and remain as hot Jupiters.
As explained below, this phenomenon is attributable to lower mass-loss rates at distant orbits and the runaway nature of mass loss.

\begin{figure}
\begin{center}
\includegraphics[scale=1.0]{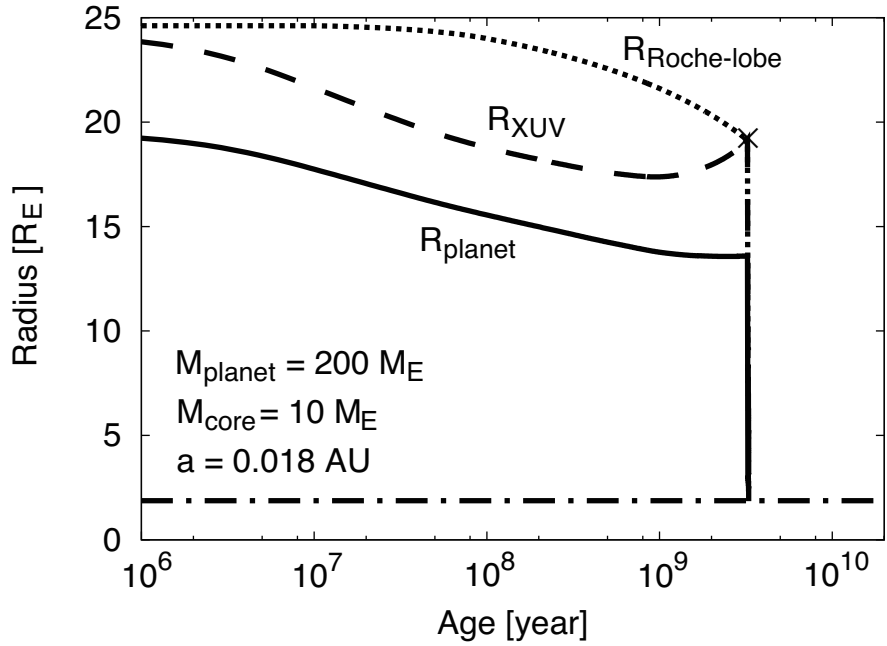}
\caption{Radial evolution of a $200\ M_{\rm Earth}$ planet as a function of time. 
The semimajor axis is $0.018\ {\rm AU}$ and the core mass is $10\ M_{\rm Earth}$. 
Shown are the planetary radius (solid line), the XUV radius (dashed line), and the Roche-lobe radius (dotted line). 
The dash-dotted line indicates the core radius. 
The cross marks Roche-lobe overflow event. \label{Evo_RRL_10Mcore300ME0015AU_time-radius}}
\end{center}
\end{figure}

\begin{figure}
\begin{center}
\includegraphics[scale=0.5]{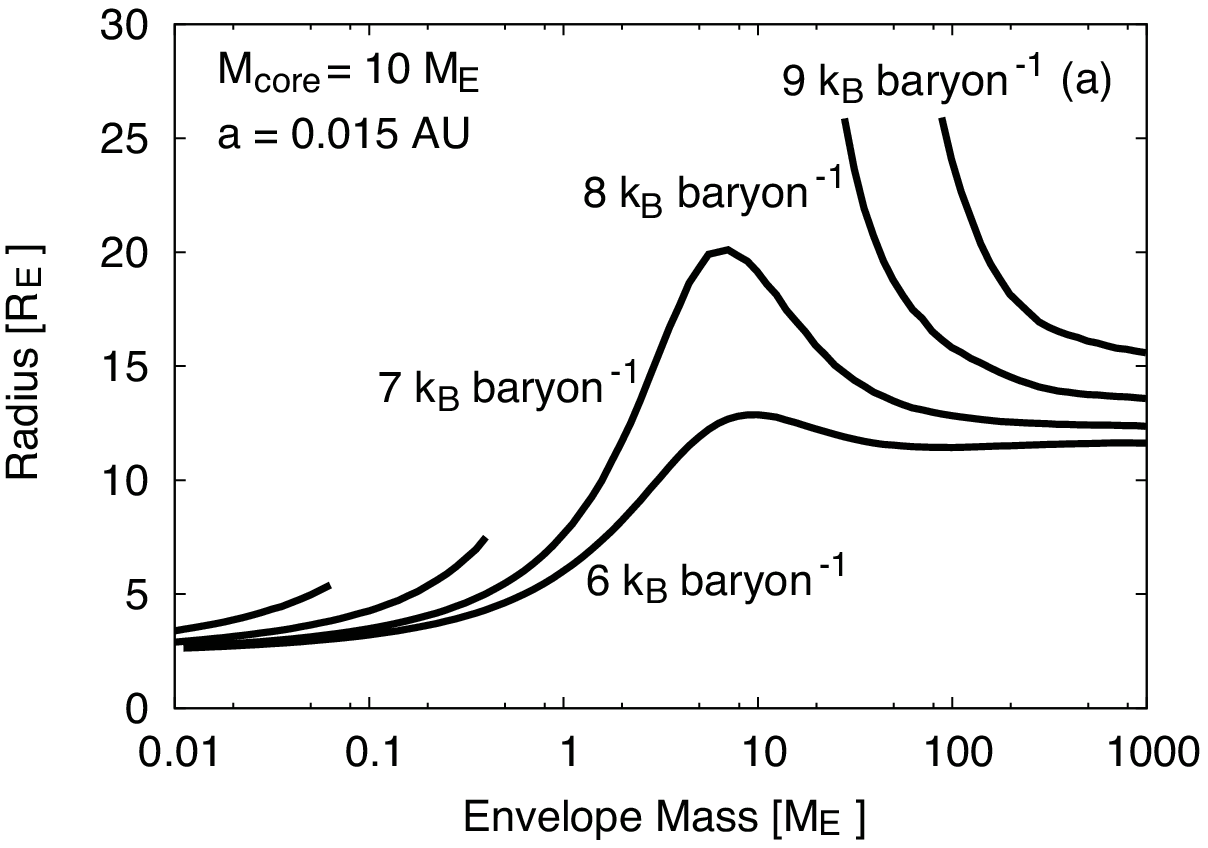}
\includegraphics[scale=0.5]{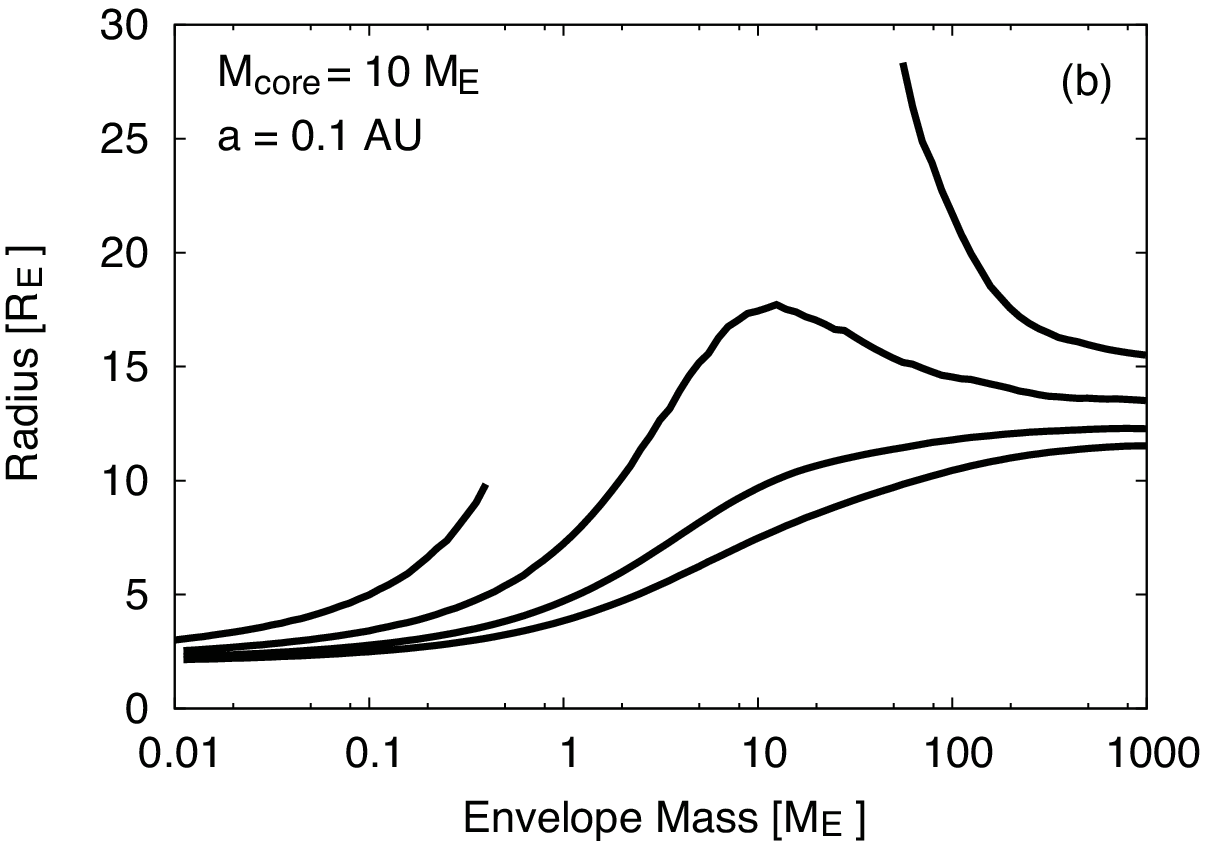}
\includegraphics[scale=0.5]{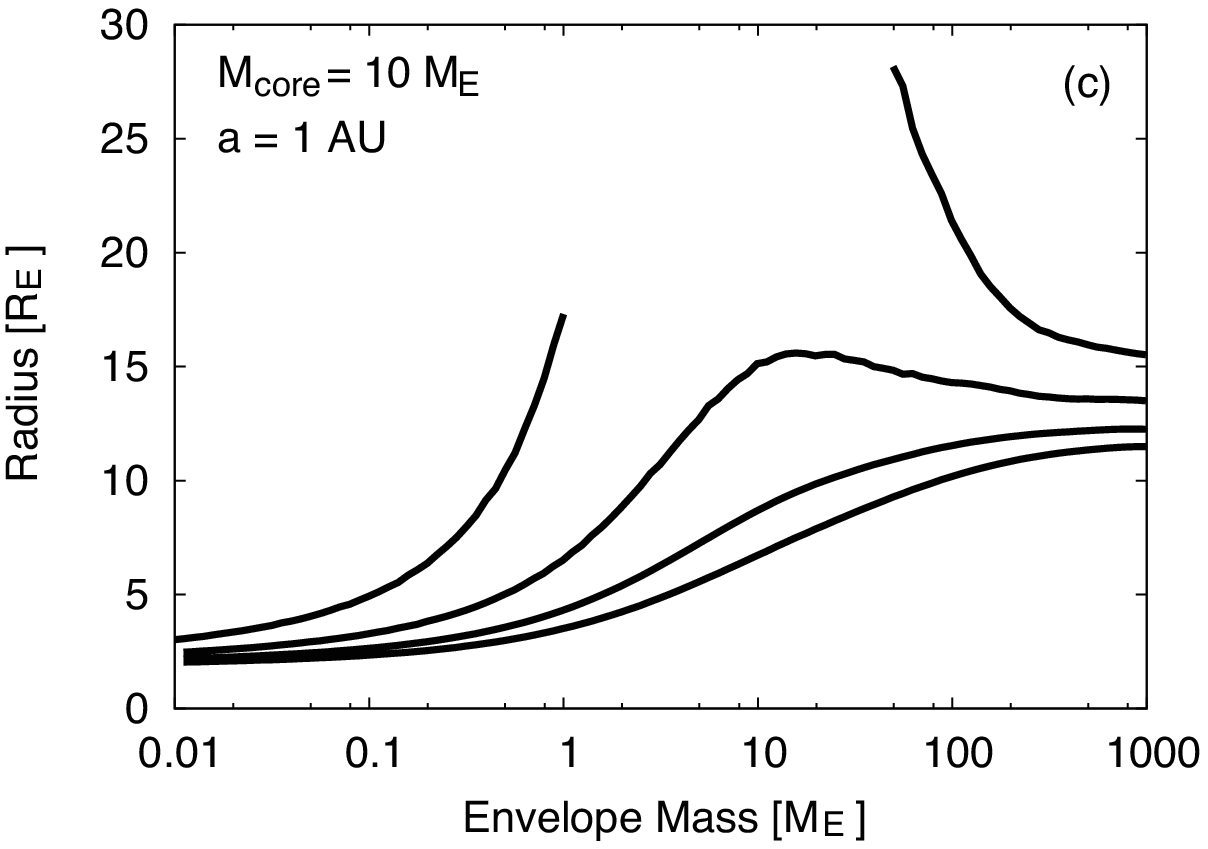}
\includegraphics[scale=0.5]{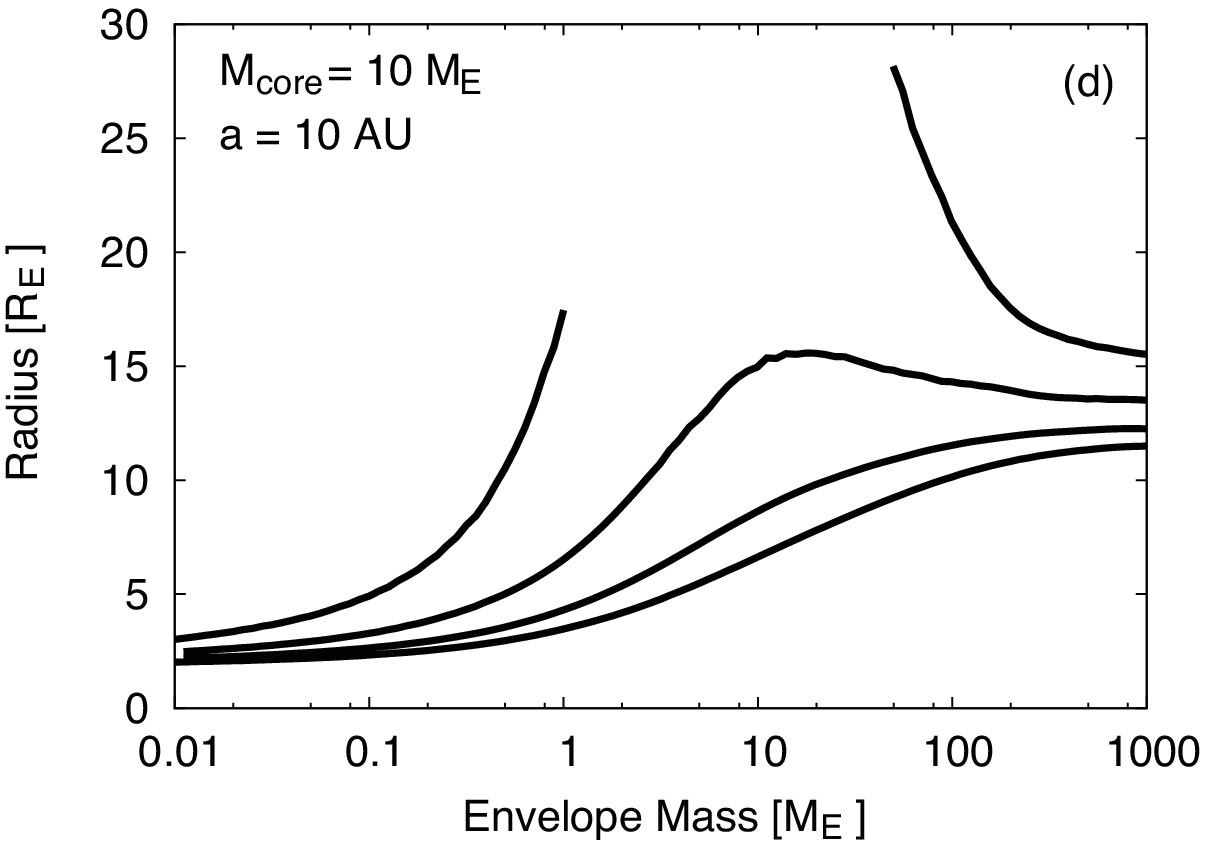}
\includegraphics[scale=0.5]{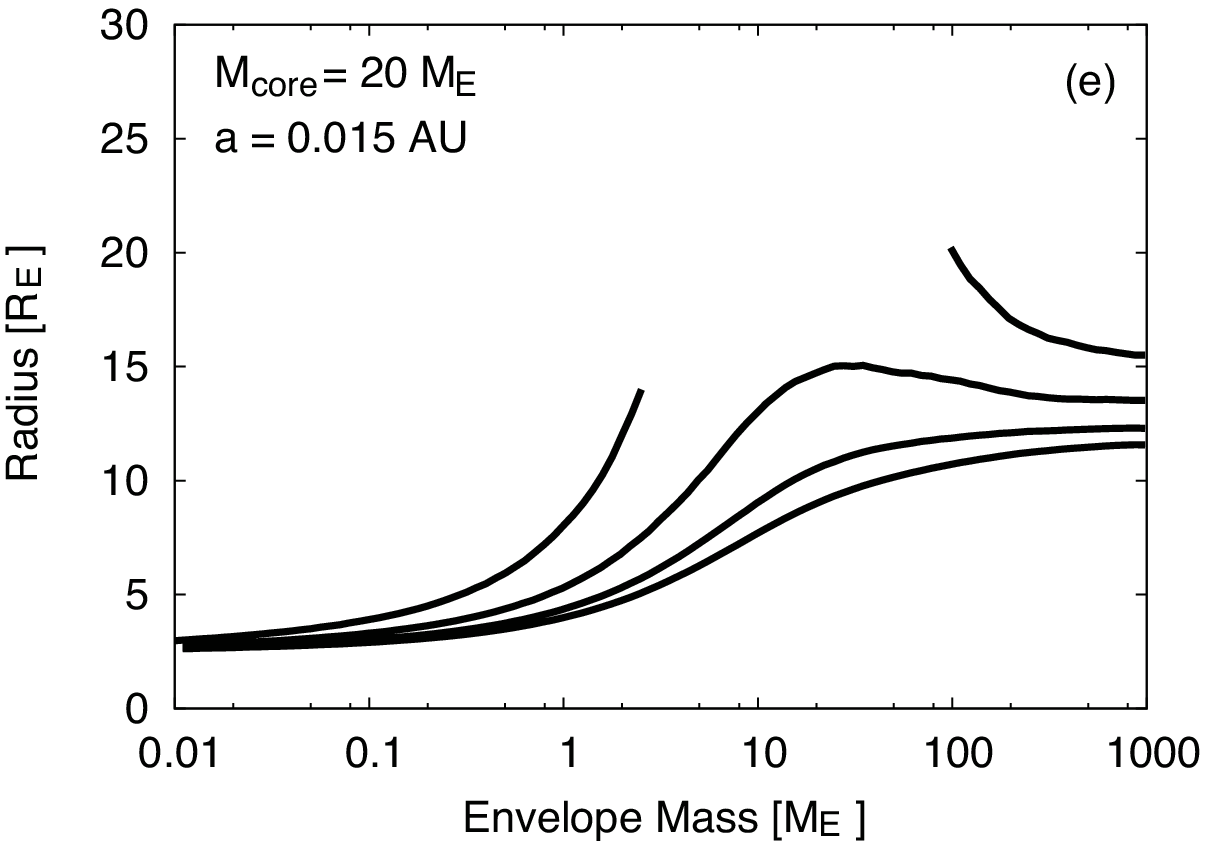}
\includegraphics[scale=0.5]{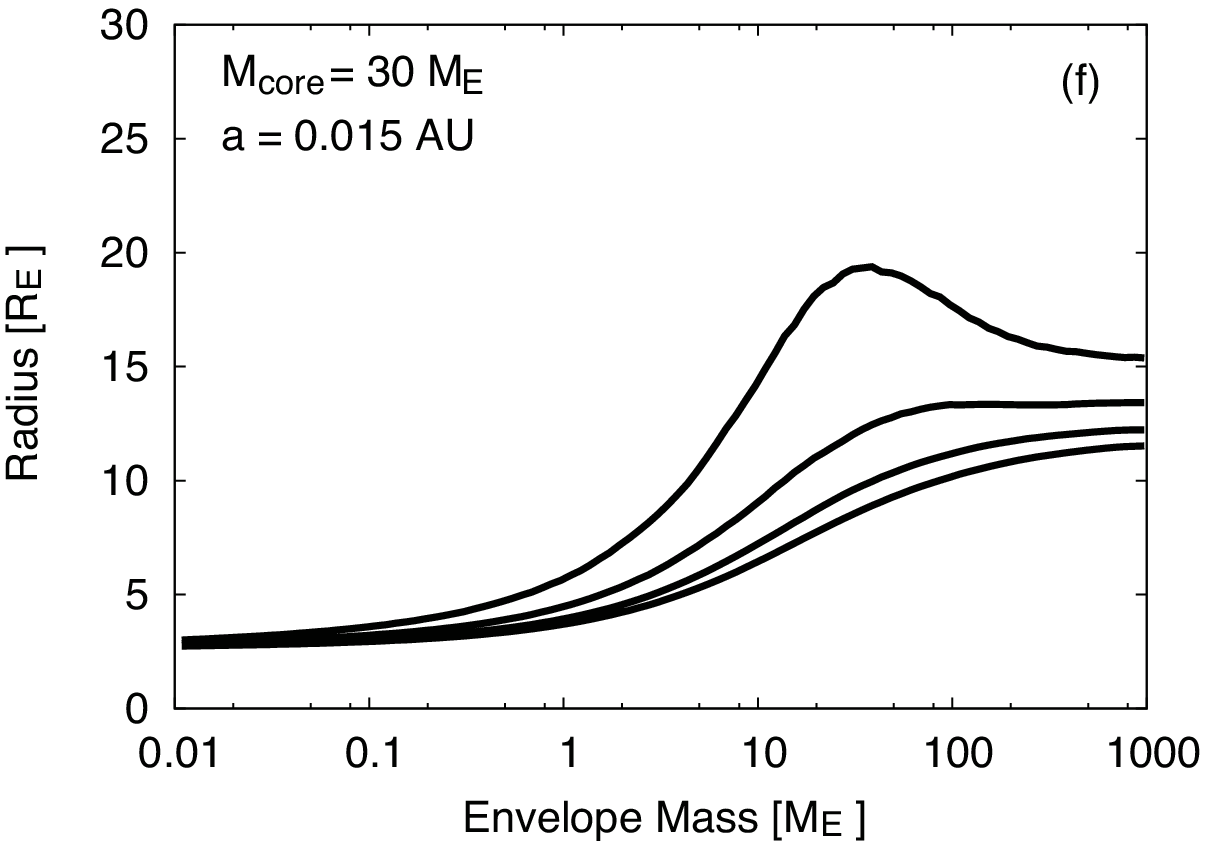}
\caption{Planetary radii as functions of envelope mass.
Core masses and semimajor axes are
(a) $10\ M_{\rm Earth}$ and $0.015\ {\rm AU}$,
(b) $10\ M_{\rm Earth}$ and $0.1\ {\rm AU}$,
(c) $10\ M_{\rm Earth}$ and $1.0\ {\rm AU}$,
(d) $10\ M_{\rm Earth}$ and $10\ {\rm AU}$,
(e) $20\ M_{\rm Earth}$ and $0.015\ {\rm AU}$,
(f) $30\ M_{\rm Earth}$ and $0.015\ {\rm AU}$.
Results (solid lines) are plotted for different entropy in the convective layer $S_{\rm convective} = 9,8,7, {\rm and}\ 6\ {\rm k_{\rm B}\ baryon^{-1}}$ (ordered from top to bottom). \label{MatmRpl_10-50Mcore_S6-9}}
\end{center}
\end{figure}

Fig. \ref{Evo_RRL_10Mcore300ME0015AU_time-radius} shows how the planetary radius, the XUV radius, and the Roche-lobe radius evolve if the planet's envelope completely evaporates. 
The Roche-lobe radius decreases with mass.
The planetary radius changes less dramatically and the XUV radius enlarges prior to Roche-lobe overflow as the gravity declines.
The expansion due to mass loss is a general property of hot planets with moderately high mass envelopes, 
as observed in Fig. \ref{MatmRpl_10-50Mcore_S6-9}, which plots the planetary radius as a function of envelope mass.
during early-stage cooling, the planetary and XUV radii are both reduced.
The cooling timescale lengthens as cooling progresses (see Figs. \ref{time-radius_10Mcore} and \ref{time-entropy_10Mcore}).
At later stages of the evolution, when the cooling timescale exceeds the timescale of mass loss, 
the planet expands as it loses mass.
This mechanism leads to runway thermal atmospheric escape followed by Roche-lobe overflow \citep{Baraffe+2004,Kurokawa+Kaltenegger2013}.
Because moderate amount of the envelope is insecurely bound, 
Roche-lobe overflow inflates the radius, leaving a thin envelope surrounding the core (Fig. \ref{MatmRpl_10-50Mcore_S6-9}).
Following Roche-lobe overflow, the radius suddenly decreases as most of its envelope is lost.

The mass-loss evolution of hot Jupiters with the same semimajor axis but different initial masses are shown in Fig. \ref{Evo_RRL_10Mcore150-350ME_0020AU}.
From slightly varying initial masses, dichotomous population evolves: 
planets with completely evaporated envelopes versus those remaining as hot Jupiters.
This dichotomy occurs because the radii of heavier planets are smaller in the Jupiter-mass regime (Fig. \ref{MatmRpl_10-50Mcore_S6-9}) and thus lose mass more slowly.
The difference between the populations is amplified throughout the evolution by runaway thermal atmospheric escape and the Roche-lobe overflow.
From this finding, we can define the \textquotedblleft minimum survival mass" as a function of semimajor axis.
The envelopes of hot Jupiters lighter than the minimum survival mass evaporate completely, 
and the planets evolve into super Earths.
Hot Jupiters exceeding this critical mass retain most of their envelopes and remain as hot Jupiters.
In the following sections, the minimum survival masses are calculated and compared with the observed exoplanet populations.

The effects of core mass and formation history on the mass-loss evolution are shown in Figs. \ref{Evo_RRL-mRRL_10-50Mcore300ME_0015-0019AU} and \ref{Evo_RRL-mRRL_10-50Mcore300ME0015AU_time-radius}.
The evolution is little affected by migration history.
Hotter planets, namely planets with higher entropy, in general have larger radii.
Because migrated planets possess low entropy during their initial mass loss, 
their radii are smaller than those of {\it in situ} planets.
Consequently, they lose mass at reduced rate; recall that rate of mass loss depends on planetary radius in both radiation-recombination and the energy-limited regimes (Fig. \ref{Fxuv-Mdot_R}).
Although these phenomena introduce slight differences in the evolutions of migrated and {\it in situ} planets, 
a cooler initial state implies a longer initial cooling timescale.
The migrated planets undergo runaway thermal escape once the mass loss timescale becomes shorter than the cooling timescale.
On the other hand, core mass exerts a greater effect in mass-loss evolution than migration history.
The relationships between envelope mass and planetary radius for different core masses at the same orbital distance ($0.015\ {\rm AU}$) are shown in Fig. \ref{MatmRpl_10-50Mcore_S6-9}.
At planetary masses $\sim 1000\ M_{\rm Earth}$, 
the radii are similar for core masses of $10\ M_{\rm Earth}$ and $30\ M_{\rm Earth}$.
Below $\sim 100\ M_{\rm Earth}$, 
the radii of lighter-core planets ($10\ M_{\rm Earth}$) increase with decreasing envelope mass, 
indicating that such planets expand as their envelopes are diminished by atmospheric escape.
As discussed above, thermal atmospheric escape and succeeding Roche-lobe overflow lead to runaway mass loss.
However, in planets with heavier cores ($30\ M_{\rm Earth}$), 
a different trend emerges.
The radius increases with decreasing envelope mass only at the highest entropy; 
otherwise, the radius monotonically decreases with reduction in envelope mass.
Thus, larger-core planets are stabilized against mass loss, 
although core mass is less important in planetary radii of heavier planets ($\sim 1000\ M_{\rm Earth}$).
Initially, the radius of a migrated planet of core mass $10\ M_{\rm Earth}$ is smaller than that of an {\it in situ} planet of core mass $30\ M_{\rm Earth}$ (see Fig. \ref{Evo_RRL-mRRL_10-50Mcore300ME0015AU_time-radius}).
However, the envelope of the former planet completely evaporates, 
while the latter planet remains as a hot Jupiter.
This result shows the importance of the radial response to the mass loss.

\subsection{Comparison with exoplanet populations}

\begin{figure}
\begin{center}
\includegraphics[scale=1.0]{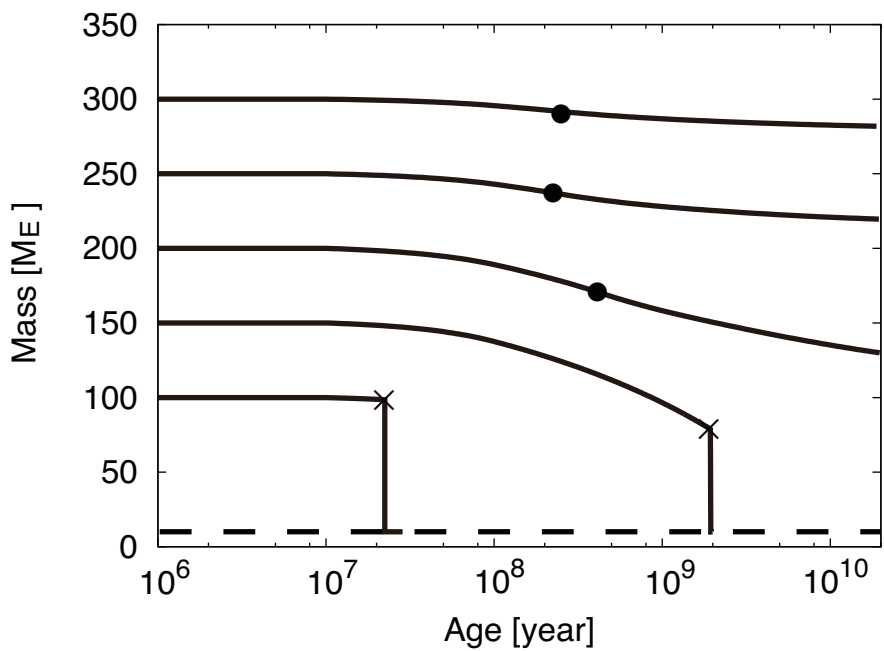}
\caption{Mass evolution of hot Jupiters as a function of time. 
All semimajor axes are $0.02\ {\rm AU}$, and core masses are $10\ M_{\rm Earth}$ (indicated by the dashed line). 
Results are plotted for planetary masses $300\ M_{\rm Earth}$, $250\ M_{\rm Earth}$, $200\ M_{\rm Earth}$, $150\ M_{\rm Earth}$, and $100\ M_{\rm Earth}$ (ordered from top to bottom). 
Filled circles indicate the changeover times of the atmospheric escape regime; from radiation-recombination limited to energy-limited. 
Crosses indicate Roche-lobe overflow events. \label{Evo_RRL_10Mcore150-350ME_0020AU}}
\end{center}
\end{figure}

By calculating mass-loss evolution at different planetary masses and semimajor axes, 
we can obtain the minimum survival mass as a function of the semimajor axis.
Within a suitable evolution time (here we assumed $10\ {\rm Gyrs}$), 
planets heavier than the critical mass are robust to mass loss, 
while lighter planets lose their entire envelope and become naked solid-core planets.
For example, the minimum survival mass of {\it in situ} planets with semimajor axis $0.02\ {\rm AU}$ and core mass $10\ M_{\rm Earth}$ is $200\ M_{\rm Earth}$ (Fig. \ref{Evo_RRL_10Mcore150-350ME_0020AU}).
Minimum survival masses are plotted as functions of  semimajor axis in Fig. \ref{aplanet-Mcritical}.
The results for different scenarios are shown.
Note that the survival mass is calculated to two significant figures.
The fiducial model comprises {\it in-situ}-formed planets of core mass $10\ M_{\rm Earth}$.
As shown in subsection 3.1, core mass exerts significant effects on the mass-loss evolution.
Because planets with heavy cores are robust against envelope loss, 
their survival masses are smaller than those of lighter-core planets.
Migrated planets, with their smaller entropy and radii, are more stable than the fiducial model.
Consequently, their survival mass is decreased, but only slightly.
For comparison, the model excluding radiation-recombination limited escape (in which the envelope is lost by energy-limited escape and Roche-lobe overflow alone), is also shown.
In general, radiation-recombination limited escape reduces the rate of mass loss, especially at larger semimajor axes.
At small semimajor axes ($< 0.03\ {\rm AU}$), 
the original and refined models yield very similar results.
Under these conditions, the rate of mass lost by radiation-recombination limited escape is enhanced by tidal effects (Fig. \ref{Fxuv-Mdot}).

Fig. \ref{observation_aplanet-Mcritical} compares the predicted minimum survival masses with the observed exoplanet populations.
Although the differences among host stellar properties are ignored in this figure, 
the majority of observed exoplanets orbit G-type stars, as assumed in our model.
In the observed population, the desert of sub-Jupiter mass planets occurs at close orbit ($\ltsim \ 0.04\ {\rm AU}$) and planetary mass of $\simeq 100\ M_{\rm Earth}$.
The minimum survival masses of planets of core mass $10\ M_{\rm Earth}$ are consistent with the observed desert.
Planets heavier than the survival mass also lose some mass, thereby moving down in the mass-distribution diagram. 
As observed in Fig. \ref{Evo_RRL_10Mcore150-350ME_0020AU}, however, slight differences in the initial mass cause large differences in final mass because mass loss is a runaway process.
Thus, we can neglect the mass-loss of planets heavier than the survival mass and directly compare the predicted line with the observed desert.
The survival masses of the heavier core models ($20\ M_{\rm Earth}, 30\ M_{\rm Earth}$) are smaller than that of the lighter core model ($10\ M_{\rm Earth}$).
The lines of heavier core models do not match the observed desert.
Therefore, our results indicate that hot Jupiters tend to have small cores ($\simeq 10\ M_{\rm Earth}$).
A core mass of $10\ M_{\rm Earth}$ is consistent with the typical mass of super Earths at the semi-major axis of the desert at $ \ltsim \ 0.04\ {\rm AU}$.
Some of these super Earths may be remnants of evaporated sub-Jupiter mass exoplanets.
Beyond $\simeq 0.04\ {\rm AU}$ semimajor axis, the mass distribution of exoplanets does not significantly depend on mass loss.

Next, we compare our results with the observed radii of exoplanets.
The observed distribution of planetary radii comprising both confirmed planets and {\it Kepler} planet candidates, is shown in Fig. \ref{observation_ap-Rp}.
Planets of sub-Jupiter radius are sparse in the range of $\simeq 3-10\ R_{\rm Earth}$ and $\ltsim \ 0.04\ {\rm AU}$.
Two confirmed planets in this cavity are known to orbit M-type stars and should thus be excluded from the discussion.
Moreover, as indicated by \citet{Beauge+Nesvorny2013}, 
all the 16 candidates found in the desert occupy single-planet systems.
The false-positive detection rates of single-planets systems are higher than those of multiple-planet systems, 
and at least 7 of the desert candidates may be false positives \citep{Beauge+Nesvorny2013}, 
which would obscure the desert boundaries.
Planets larger than $10\ R_{\rm Earth}$ and smaller than $3\ R_{\rm Earth}$ can be regarded as hot Jupiters and super Earths, respectively; 
the latter are devoid of massive envelopes.
Assuming a core mass of $10\ M_{\rm Earth}$ (consistent with the sub-Jupiter mass desert), 
radii of $3-10\ R_{\rm Earth}$ correspond to envelope masses of $< 10\ M_{\rm Earth}$ at close orbits ( $<\ 0.04\ {\rm AU}$), 
as shown in Fig. \ref{MatmRpl_10-50Mcore_S6-9}.
However, because the masses of these planets are below the minimum survival masses, 
their envelopes could be evaporated.
Therefore, our results suggest that mass loss may be responsible for the dearth of planets with sub-Jupiter radii.

\section{Discussion}

\subsection{Comparison with previous studies}

\begin{figure}
\begin{center}
\includegraphics[scale=1.0]{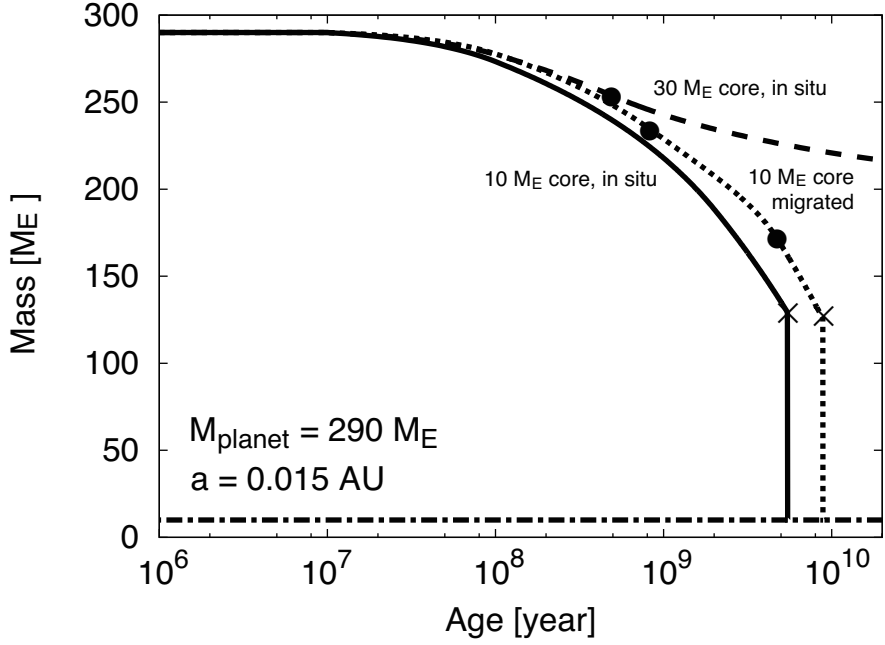}
\caption{Mass evolution of $290\ M_{\rm Earth}$ planets as a function of time. 
All semi-major axes are $0.015\ {\rm AU}$. 
The planets formed {\it in situ} have core masses $10\ M_{\rm Earth}$ (solid line) and $30\ M_{\rm Earth}$ (dashed line), 
The core mass of the migrated planet is $10\ M_{\rm Earth}$ (dotted line). 
The dash-dotted line indicates the mass of the $10\ M_{\rm Earth}$ core. 
Filled circles indicate the changeover times of the atmospheric escape regime; from radiation-recombination limited to energy-limited. 
The second circle in the migrated case marks the transition from the energy-limited regime to the radiation-recombination limited regime. 
Crosses indicate Roche-lobe overflow events. \label{Evo_RRL-mRRL_10-50Mcore300ME_0015-0019AU}}
\end{center}
\end{figure}

\begin{figure}
\begin{center}
\includegraphics[scale=1.0]{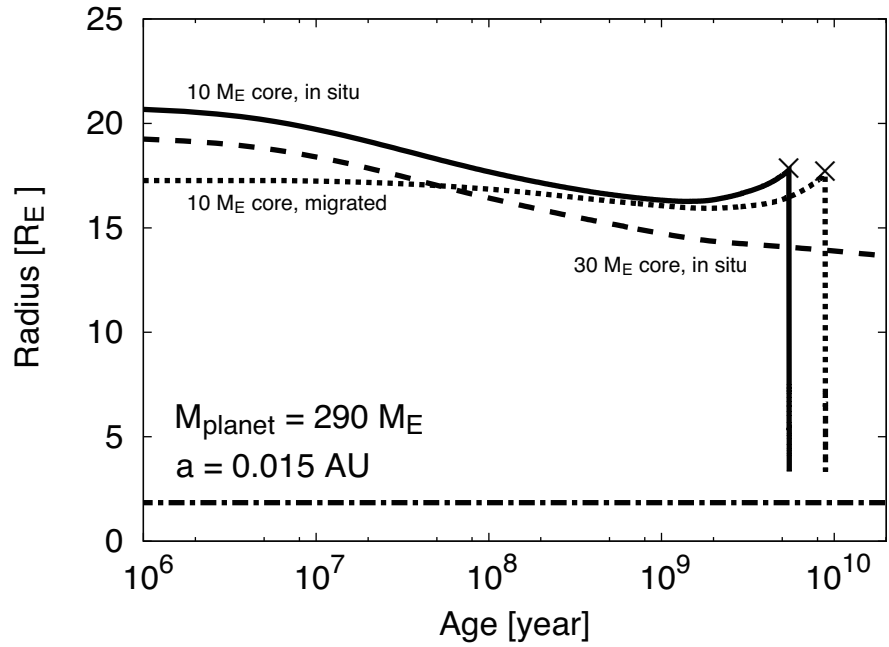}
\caption{Evolution of the XUV radii $R_{\rm XUV}$ of Jupiter-mass ($290\ M_{\rm Earth}$) planets as functions of time. 
The settings are described in the caption of Fig. \ref{Evo_RRL-mRRL_10-50Mcore300ME_0015-0019AU}.
The dash-dotted line indicates the radius of a $10\ M_{\rm Earth}$ core.
\label{Evo_RRL-mRRL_10-50Mcore300ME0015AU_time-radius}}
\end{center}
\end{figure}

\begin{figure}
\begin{center}
\includegraphics[scale=1.0]{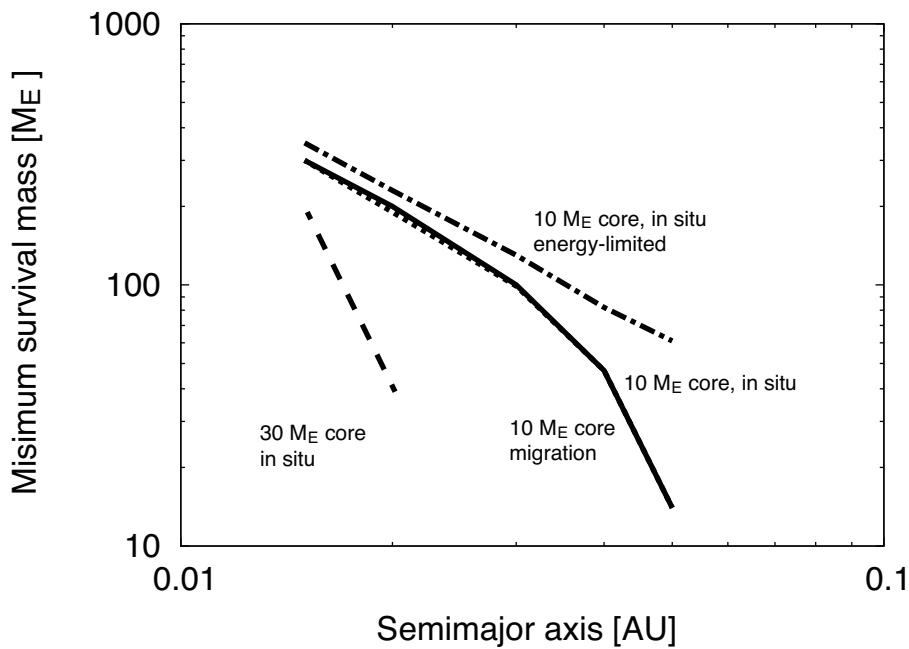}
\caption{Minimum survival mass as a function of semimajor axis. 
A solid line is the result of planets which have $10\ M_{\rm Earth}$ cores and are formed {\it in situ}, a dash-dotted line is the same with the solid line except for the assumption of the energy-limited escape, a dotted line is the result of planets which have $10\ M_{\rm Earth}$ cores and are migrated, and a dashed line is the result of planets which have $30\ M_{\rm Earth}$ cores and are formed in situ, respectively. 
Note that the solid and the dotted lines ($10\ M_{\rm Earth}$ core, {\it in situ} formation and migrated) are almost overlapped.
\label{aplanet-Mcritical}}
\end{center}
\end{figure}

\begin{figure}
\begin{center}
\includegraphics[scale=1.0]{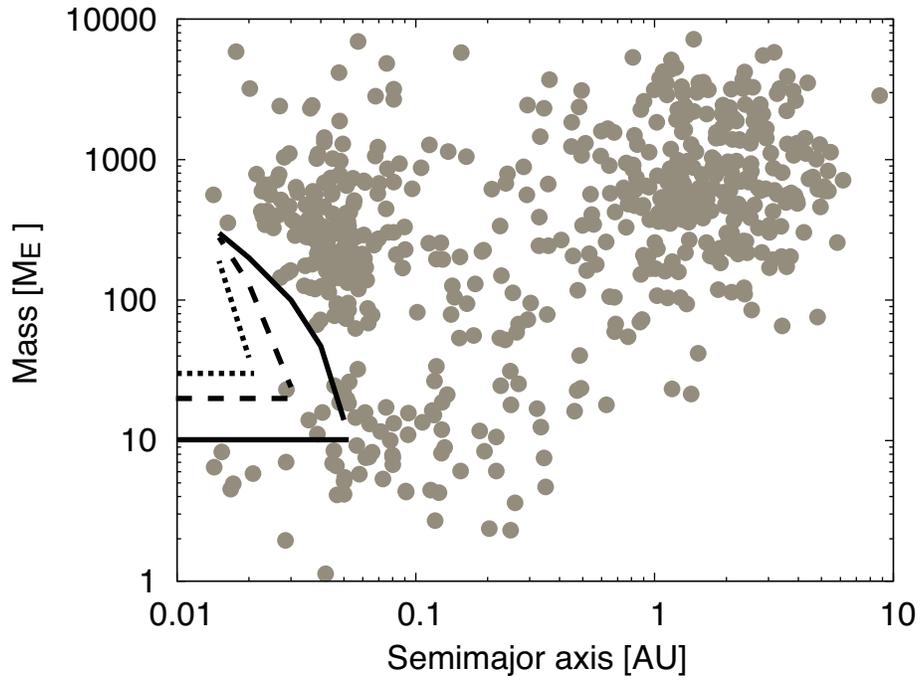}
\caption{Comparison of minimum survival mass with observed population of exoplanets. Data points are taken from exoplanet.org on August 27th, 2013. Lines are minimum survival masses of planets formed {\it in situ} with core masses of $10\ M_{\rm Earth}$ (solid line), $20\ M_{\rm Earth}$ (dashed line), and $30\ M_{\rm Earth}$ (dotted line).
Corresponding core masses are shown by horizontal lines.
\label{observation_aplanet-Mcritical}}
\end{center}
\end{figure}

\begin{figure}
\begin{center}
\includegraphics[scale=1.0]{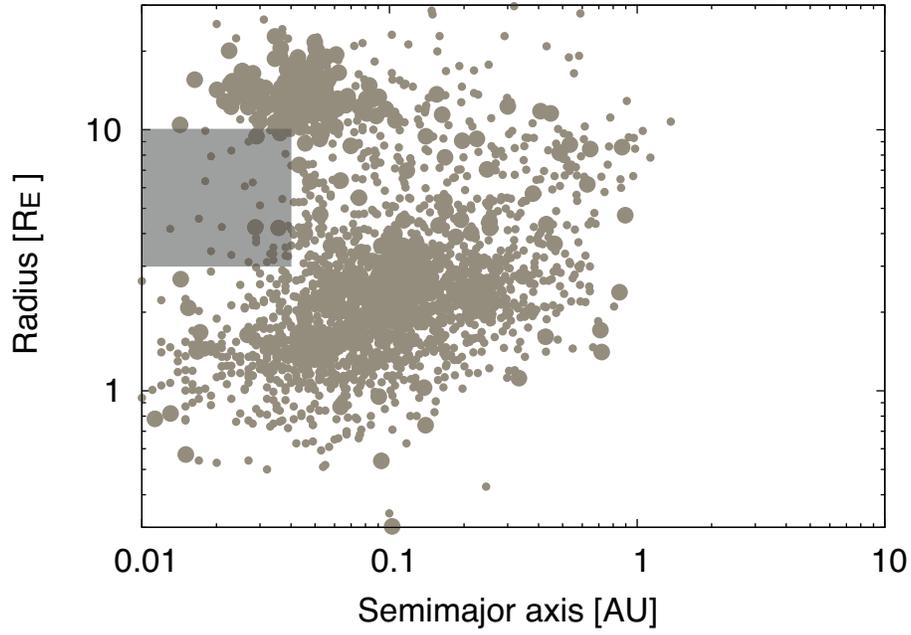}
\caption{Observed radius and semimajor axis of confirmed exoplanets (larger points) and {\it Kepler} planet candidates (smaller points). 
Data points of confirmed exoplanets are taken from exoplanet.org on August 27th, 2013.
Data points of {\it Kepler} planet candidates are taken from {\it The Mikulski Archive for Space Telescopes (MAST)}.
The desert (delineated by the gray square) is discussed in the text.
Two of the data points in the desert (GJ 436 b and GJ 3470 b) orbit M-type stars and are therefore excluded from our analysis of hot Jupiters orbiting G-type stars.
\label{observation_ap-Rp}}
\end{center}
\end{figure}

Recently mass-loss evolution of close-in exoplanets was studied by \citet{Owen+Wu2013} and \citet{Lopez+Fortney2013}.
The model of \citet{Lopez+Fortney2013} is similar with ours but the radiation-recombination limited escape is not included.
The model of \citet{Owen+Wu2013} used the rate of mass loss obtained by their hydrodynamic simulation and the XUV model is different.
Though these studies mainly concerned low mass planets, 
a few examples of heavier planets were studied.
\citet{Owen+Wu2013} showed evolution of a Jupiter mass ($318\ M_{\rm Earth}$) planet having a $15\ M_{\rm Earth}$ core at $0.025\ {\rm AU}$ in their Fig. 2 and concluded that the mass loss is enough small compared to the total mass ($\sim 0.5\ \%$ of the total mass is lost after $10\ {\rm Gyr}$). 
The heaviest planet of \citet{Lopez+Fortney2013} calculations is a $320\ M_{\rm Earth}$ planet having a $64\ M_{\rm Earth}$ core at $\sim 0.033\ {\rm AU}$ (converted from the assumed incident flux $1000$ times larger than that Earth receives) in their Fig.2, 
which does not lose significant mass.
In our fiducial model (having a $10\ M_{\rm Earth}$ core and including the radiation-recombination limited escape), 
a high mass planet ($290\ M_{\rm Earth}$, nearly a Jupiter mass) completely evaporates at a quite close-in orbit ($0.015\ {\rm AU}$, our Fig. \ref{Evo_RRL-mRRL_10-50Mcore300ME_0015-0019AU}). 

We compare our results with these examples of high mass planets in \citet{Owen+Wu2013} and \citet{Lopez+Fortney2013}, and conclude that the main cause for the difference of results is the difference of the assumed separation from the host star. 
The amount of mass loss strongly depends on the assumed separation. 
Planets orbiting a slightly far separation hardly evaporate as shown in our Figs. \ref{Evo_RRL_10Mcore300ME_0015-0019AU} and \ref{observation_aplanet-Mcritical}. 
This is due to the runaway property of the mass-loss evolution discussed in this study. 
We calculated the evolution of a Jupiter mass planet having a $15\ M_{\rm Earth}$ core at $0.025\ {\rm AU}$ (not shown as a figure), which is the same as Fig. 2 of \citet{Owen+Wu2013}. 
The planet lost only $1.6\ \%$ of the total mass ($\sim 5\ M_{\rm Earth}$) after $10\ {\rm Gyr}$, 
which is three times larger than the result of \citet{Owen+Wu2013} and does not differ by an order of magnitude. 
The slight difference of the lost mass is caused by different mass loss models. 
We used a semianalytical model including the recombination limited escape which corresponds to \textquotedblleft EUV-driven" regime in \citet{Owen+Wu2013}. 
\citet{Owen+Wu2013} used the rate of mass loss obtained by their hydrodynamic simulation and the mass loss is mainly in \textquotedblleft X-ray-driven" regime. 
The difference of XUV models also affects the results. 
The difference from \citet{Lopez+Fortney2013} is due to the higher core mass ($64\ M_{\rm Earth}$) of their $320\ M_{\rm Earth}$ planet as well as their large separation.
The core mass strongly affects the mass-loss evolution as shown in our Figs. \ref{Evo_RRL-mRRL_10-50Mcore300ME_0015-0019AU} and \ref{aplanet-Mcritical}. 

\subsection{Validity of model}

\begin{figure}
\begin{center}
\includegraphics[scale=1.0]{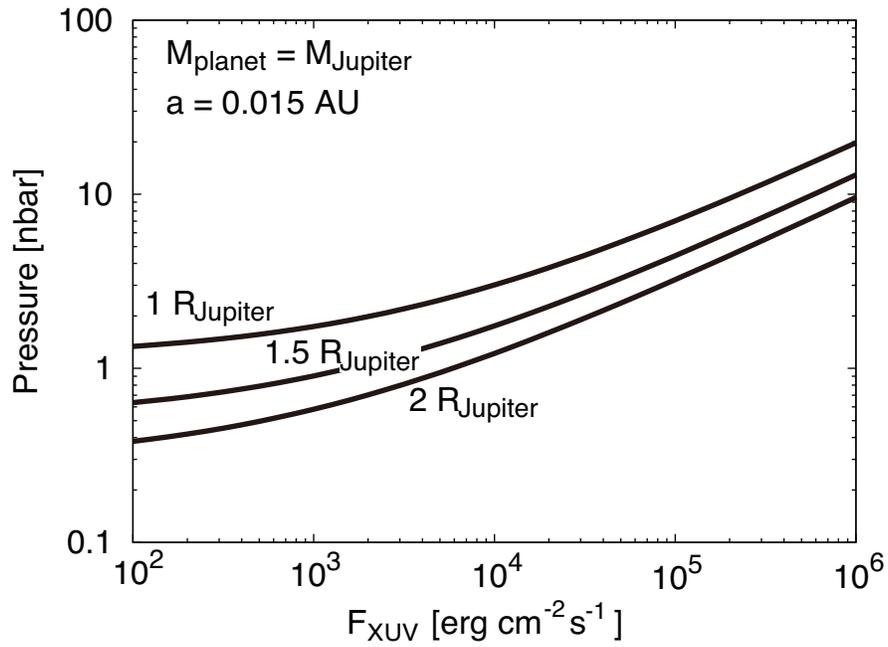}
\caption{
Pressure at the ionization front in the framework of our semianalytical model obtained by using our Eqs. \ref{rho+_base} and \ref{rho0_base}. 
Results are plotted for a Jupiter mass planet at $0.015\ {\rm AU}$ having $1$, $1.5$, $2\ R_{\rm Jupiter}$, respectively.
\label{FXUV-Pressure}}
\end{center}
\end{figure}

\begin{figure}
\begin{center}
\includegraphics[scale=1.0]{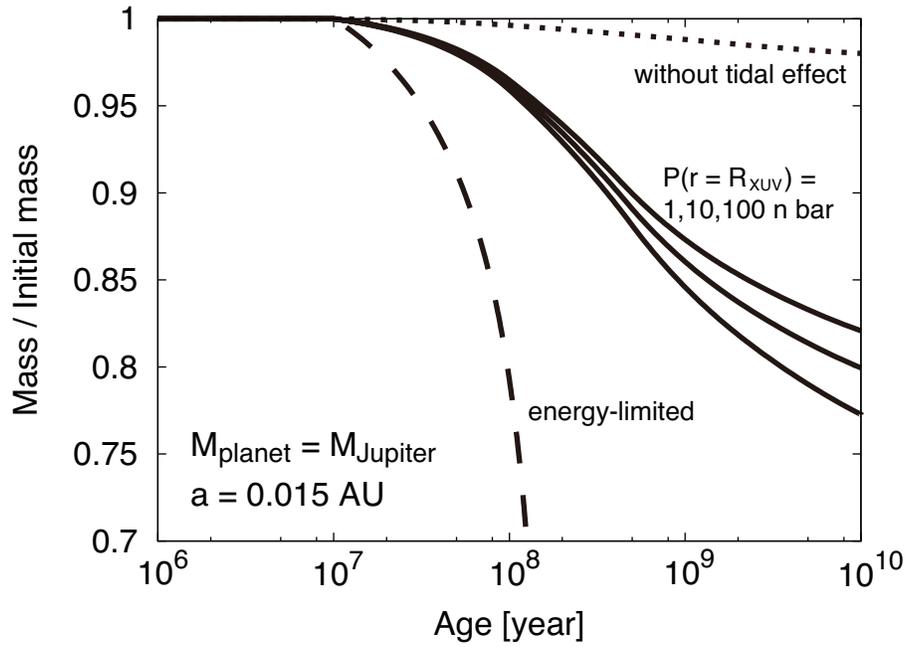}
\caption{
Mass evolution of Jupiter mass planets. 
The semimajor axis is 0.015 AU. 
Shown are results obtained by using models with artificially changed pressure to evaluate $R_{\rm XUV}$ (solid lines, $1$, $10$, $100\ {\rm nbar}$, ordered from bottom to top), a model without radiation-recombination limited regime (a dashed line), and a model without tidal effects (a dotted line).
\label{Time-Mass_OW2013}}
\end{center}
\end{figure}

We calculated the XUV radius ${\rm R_{XUV}}$ by assuming the pressure of $1\ {\rm nbar}$ as a crude estimate.
The pressure at the ionization front (which is identical with the XUV radius) can be obtained in the framework of our semianalytical model by using our Eqs. \ref{rho+_base} and \ref{rho0_base} (Fig. \ref{FXUV-Pressure}). 
The pressure at the ionization front increases as a function of XUV flux and becomes $\sim 10\ {\rm nbar}$ at higher XUV level. 
This increase of the pressure (namely, the increase of the number density) can be found in \citet{Murray-Clay+2009} and \citet{Owen+Jackson2012}. 
We evaluate the error caused by our assumption for the pressure at the XUV radius in Fig. \ref{Time-Mass_OW2013}.
The amount of mass loss changes only a few percent when we artificially change the pressure at ${\rm R_{XUV}}$. 
Mass loss is more affected by the presence of the radiation-recombination limited escape. 
The rate of mass loss in the radiation-recombination limited regime is enhanced by the tidal effect, but the amount of mass loss is reduced than that obtained without the radiation-recombination limited regime (the energy-limited regime only).

Pre-main-sequence (PMS) stars initially have radii $\simeq 3$ times larger than the solar radius \citep[the case of 1 solar mass,][]{Palla+Stahler1993}, which corresponds to $\simeq 0.014\ {\rm AU}$, and shrink through time to evolve into main-sequence (MS) stars.
Innermost cases of observed planets and our model ($0.015\ {\rm AU}$, Fig.\ref{observation_aplanet-Mcritical}) are close to being inside the radii of early PMS stars.
Planets at this innermost orbit might be engulfed at the early PMS stage. 
The observed innermost planets are possibly migrated after the stage. 
Because the survival masses of the migrated planets are almost the same as those of the planets formed {\it in situ} (Fig. \ref{aplanet-Mcritical}), 
the planet engulfment and migration do not change our discussion on planet evaporation.

Inner cavities of protoplanetary disks occur at $\simeq 0.03-0.04\ {\rm AU}$ \citep[][and references therein]{Najita+2007}.
Planets which migrated inside the cavity are irradiated by stellar XUV radiation at the PMS stage before dissipation of the protoplanetary disks.
Observations show that the X-ray luminosities of PMS stars are $\sim 10^{30}\ {\rm erg\ s^{-1}}$ on average \citep{Gudel+2007}, which are similar with MS stars in saturation phase \citep{Jackson+2012}.
If the X-ray luminosity scales with the bolometric luminosity as suggested by observations \citep{Gudel+2007}, 
early PMS stars, having a few times larger bolometric luminosity than MS stars within a few million years \citep{Palla+Stahler1993}, might emit a few times larger X-ray luminosity.
Assuming the same dependence for XUV luminosity, 
our model, which assumed that the mass loss starts after the disk dissipation, provides the minimum estimate for the survival mass.
Because the cumulative XUV energy in the neglected $10^7\ {\rm years}$ is smaller than the XUV energy in saturation phase ($10^8\ {\rm years}$), 
however, the effect would be small.

\subsection{Effects of uncertainties}

\begin{figure}
\begin{center}
\includegraphics[scale=1.0]{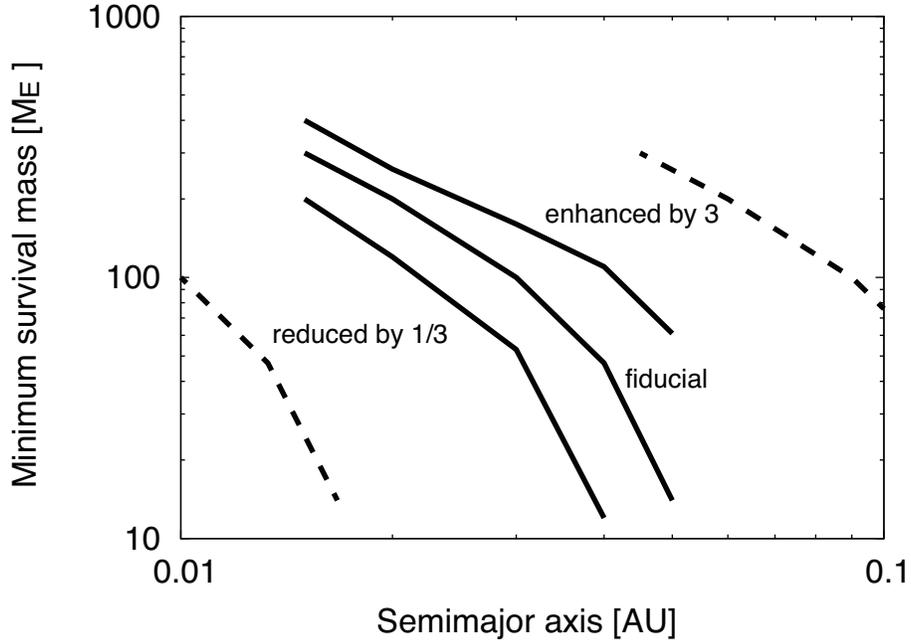}
\caption{
Minimum survival mass as a function of semimajor axis. 
The middle solid line plots the result of planets formed {\it in situ} with core mass $10\ M_{\rm Earth}$.
The left and right solid lines plot identical situations, but with the rate of mass loss reduced by a factor of $3$ and enhanced by a factor of $3$, respectively.
Dotted lines are shifted from the middle solid line by factors of $1/3$ and $3$ in semimajor axis, that are expected only from the changes of mass-loss rate in the radiation-recombination limited escape.
\label{aplanet-Mcritical_t3d3}}
\end{center}
\end{figure}

\begin{figure}
\begin{center}
\includegraphics[scale=1.0]{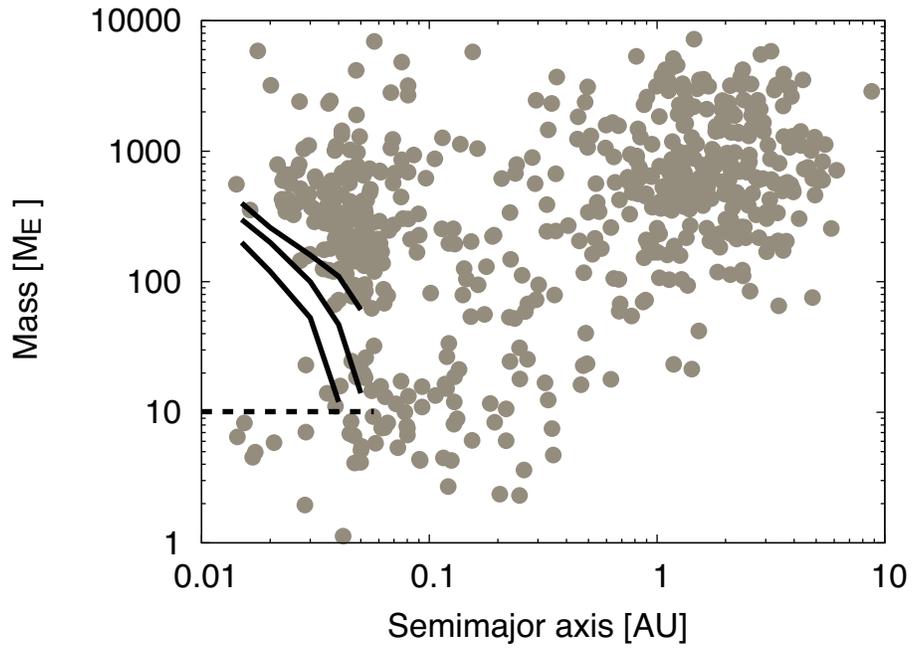}
\caption{Comparison of minimum survival mass with observed population of exoplanets. 
Minimum survival masses are obtained with the reduced and enhanced rates of mass loss presented in Fig. \ref{aplanet-Mcritical_t3d3}.
Data points are those of Fig. \ref{observation_aplanet-Mcritical}.
The dashed line represents the core mass.
\label{observation_aplanet-Mcritical_sensitivity}}
\end{center}
\end{figure}

Stellar X-ray luminosities vary by an order of magnitude even among similarly aged stars in a cluster \citep{Jackson+2012}.
Though difficult to ascertain by observations, the stellar EUV luminosity is expected to be similarly diverse \citep{Sanz-Forcada+2011}.
In addition, the efficiency of thermal atmospheric escape is uncertain by a factor of $\simeq 3$ \citep{Leitzinger+2011,Lopez+2012}.
To elucidate the effects of such diversity and uncertainty on our results, 
we plot the minimum survival masses as functions of semimajor axis for arbitrarily altered rates of mass loss (see Fig. \ref{aplanet-Mcritical_t3d3}).
These results show an insensitivity on the assumed mass-loss rates compared to that expected only from changes to compensate the mass-loss rates by XUV flux.
This insensitivity arises from the effect of stellar irradiation on planetary radius and partially from tidal effects on mass-loss rates.
The minimum survival masses obtained by suppressing and accelerating the mass loss are compared with the observed exoplanets distribution in  Fig. \ref{observation_aplanet-Mcritical_sensitivity}.
Comparing Figs. \ref{observation_aplanet-Mcritical} and \ref{observation_aplanet-Mcritical_sensitivity}, 
we observed that varying the rate of mass loss exerts smaller effect than that exerted by varying the core mass.
The sub-Jupiter desert is best fitted  by our fiducial model (defined in previous sections).
The reduced model partially reproduces the desert, 
but its survival masses locate below the upper boundary of the desert.
The survival masses of the enhanced model locate slightly above the desert; 
thus, heavier cores favor the formation of the sub-Jupiter desert by evaporation.

\begin{figure}
\begin{center}
\includegraphics[scale=1.0]{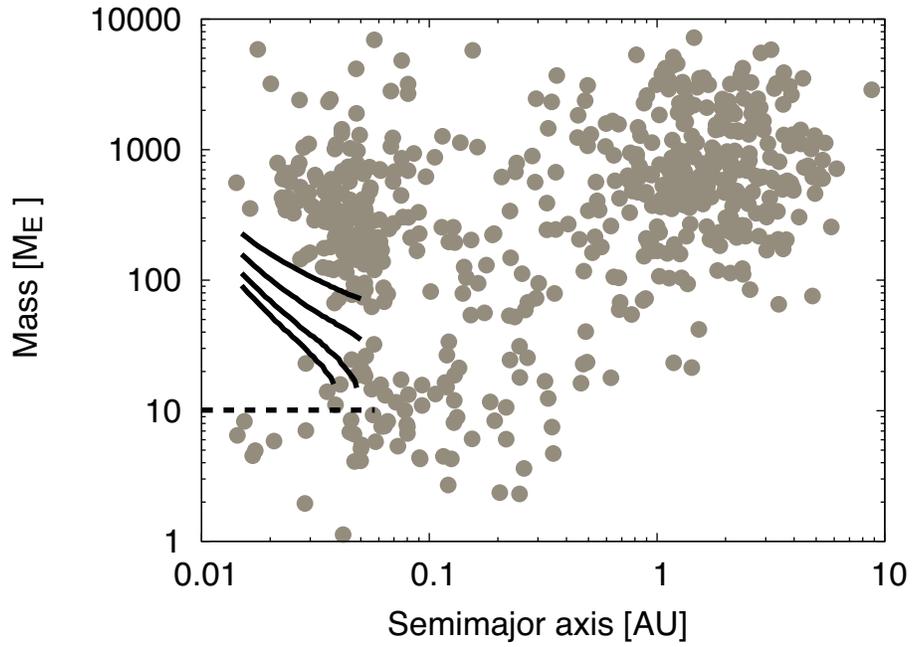}
\caption{
Comparison of critical mass for Roche-lobe overflow with observed population of exoplanets.
The critical mass is defined by $R_{\rm XUV} = R_{\rm rl}$.
Results (solid lines) are plotted for different entropy in the convective layer $S_{\rm convective} = 9,8,7, {\rm and}\ 6\ {\rm k_{\rm B}\ baryon^{-1}}$ (ordered from top to bottom).
The data points are those of Fig. \ref{observation_aplanet-Mcritical}.
The dashed line represents the core mass.
\label{AU_criticalmass_S6789}}
\end{center}
\end{figure}

In the model, 
if the $R_{\rm XUV}$ of a planet exceeded $R_{\rm rl}$ at the initial entropy of $9.2\ k_{\rm B}\ {\rm baryon}^{-1}$, 
$R_{\rm XUV}$ was reset to $R_{\rm rl}$.
This step was implemened to ensure stability of planets to Roche-lobe overflow and to elucidate its effects following thermal atmospheric escape.
Indeed, migration can induce Roche-lobe overflow in the absence of thermal atmospheric escape \citep{Trilling+1998,Gu+2003,Nayakshin+Lodato2012}.
The critical planetary masses for migration-driven Roche-lobe overflow, 
as a function of semimajor axis at different entropies are shown in Fig. \ref{AU_criticalmass_S6789}.
The entropy of planets migrating prior to disk dissipation is $\simeq 8-9\ k_{\rm B}\ {\rm baryon}^{-1}$.
Although migration-driven Roche-lobe overflow creates a desert of sub-Jupiters at inner orbits, 
the desert is smaller than observed, 
and the slopes of lines of critical mass are flatter than the upper boundary of the sub-Jupiter desert.
Therefore, mass loss processes require thermal atmospheric escape to produce the observed desert patterns.

\subsection{Implications for planet formation theory}

To gain insights into planetary formation, 
we compared the results of planetary formation models with observations.
Population syntheses of planetary formation have been previously researched; 
for example, in a series of papers by Ida \& Lin and Mordasini et al.
According to these studies, 
planets of sub-Jupiter mass are scarce at $\ltsim \ 1\ {\rm AU}$ because Jupiter-like planets form at large distances ($> \ 1\ {\rm AU}$) from their host star and migrate inward.
These results are inconsistent with the observed planet populations, 
in which no desert exists at $\ltsim \ 1\ {\rm AU}$ \citep{Howard+2010,Howard+2012}.
As discussed earlier, 
the observed sub-Jupiter desert is compacted into $\ltsim \ 0.04\ {\rm AU}$.
Thus, another mechanism must be responsible for the observed desert.
As we have shown, the desert is consistent with evaporation of the envelopes of sub-Jupiter planets orbiting close to their host star.
Another likely mechanism, proposed by \citet{Benitez-Llambay+2011}, 
is planetary migration and trapping near the inner edge of the disk.
Calculating the interaction between the planet and the disk gas, 
they showed that this mechanism reproduces the mass-period distribution of inner-orbit exoplanets $<\ 0.04\ {\rm AU}$.
In their calculation, smaller migratory planets are trapped at the inner edge, 
while heavier planets penetrate the inner edge and are trapped at the orbit whose mean motion resonates at 2:1 with the inner edge.
The critical mass is $\simeq 1$ Jupiters.
However, this mechanism predicts a desert of both sub-Jupiters and super Earths.
In real planetary populations, many super Earths compatible to hot Jupiters orbit their host stars within $0.04\ {\rm AU}$ \citep{Howard+2012}.
A third possible mechanism is that most closely orbiting ($<\ 0.04\ {\rm AU}$) exoplanets result from tidal trapping of planets with eccentric orbits.
\citet{Beauge+Nesvorny2013} proposed that inefficient trapping of smaller planets can explain the observed desert because the gas-poor composition of such planets precludes efficient dynamical  tides.
This scenario has yet to be numerically tested, 
but is expected to also predict a desert of closely orbiting super Earths, 
even if it successfully reproduces the desert of sub-Jupiters.
Therefore, neither of these alternative mechanisms can directly explain why the sub-Jupiter desert coexists with an abundance of  closely-orbiting super Earths.
One possible solution is that a fraction of inwardly migrating planets halts at  $\ltsim \ 0.04\ {\rm AU}$, 
possibly in the inner cavity of the protoplanetary disk, 
during planetary formation.
Following formation and migration, their envelopes evaporate to create the desert of sub-Jupiters.

To produce the sub-Jupiter desert by evaporation processes, 
constraints must be placed on the planetary formation scenario.
As discussed above, hot Jupiters with large cores ($\simeq 20-30\ M_{\rm Earth}$) largely retain their envelopes.
The model is consistent with observation if a core mass of $10\ M_{\rm Earth}$ is assumed (Fig. \ref{observation_aplanet-Mcritical}).
Thus, our results suggest that the cores of hot Jupiters are typically small.
This scenario requires efficient gas accretion onto the core during planet formation.
The migration history of interactions with disk gas everts minor influence on our results; 
hence, the migration of hot Jupiter by interactions with disk gas is not apparently constrained.
However, if interplanetary interactions induce the migration of multiple sub-Jupiters after $\sim 0.1-1\ {\rm Gyr}$, 
these planets should be detected in the desert because they would have received no intense XUV radiation from their host stars during their youth.
This implies that the population of inner exoplanets is not dominated by such later migrating planets.
In the radius distribution of likely {\it Kepler} planets, 
the number of super Earths exceeds that of hot Jupiters by $30$ times at $\ltsim \ 0.25\ {\rm AU}$; 
however, the subpopulations are approximately equal at $\ltsim \ 0.04\ {\rm AU}$ \citep{Howard+2012}.
A fraction of these super Earths might be remnants of evaporated hot Jupiters. 
Super-Earths occupy multiple-planet systems at a higher ratio (relative to single planet systems) than hot Jupiters \citep[see][]{Beauge+Nesvorny2013}.
Thus, the fraction of evaporated remnants might be higher for super Earths occupying single-planet systems than those occupying multiple systems.

\section{Summary and conclusions}

The exoplanet population is characterized by a desert of sub-Jupiter planets at $\ltsim 0.04\ {\rm AU}$.
We developed a numerical model that calculates the mass loss and thermal evolution of planets, 
accounting for XUV-driven thermal atmospheric escape.
Atmosphere is lost by both energy-limited escape and radiation-recombination limited escape.
Further loss occurs via the Roche-lobe overflow that is induced by and follows atmospheric escape.
We showed that the runaway property of the mass loss leads to a dichotomous population in which heavier planets remain as hot Jupiters, while smaller planets completely evaporate leaving naked core.
The results strongly depend on the core mass and weakly on migration history.
The observed sub-Jupiter desert in both mass and radius distributions a was successfully reproduced by modeling evaporation of sub-Jupiters with small cores ($10\ M_{\rm Earth}$).
Comparing our results with other possible explanations for the desert and considering the abundance of inner-orbit super Earths, 
we conclude that evaporation most likely explains the sub-Jupiter desert.

\clearpage
\acknowledgments

The authors acknowledge Tristan Guillot for fruitful discussions and the referee for comments that improved the draft.
This work is supported from Global COE program “From the Earth to Earths” and 
Grants-in-Aid from the Ministry of Education, Culture, Sports, Science and Technology (MEXT) of Japan (23244027 and 24340102).

\clearpage

\clearpage




\end{document}